\documentclass[useAMS, usenatbib]{mnras}
\usepackage{comment}
\usepackage{graphicx}
\usepackage{multirow}
\usepackage{multicol}
\usepackage{bmpsize}
\usepackage{varioref}
\usepackage{colortbl}
\usepackage{comment}
\usepackage{soul}
\usepackage{lineno}
\usepackage{import}
\bibpunct{(}{)}{;}{a}{}{,}
\usepackage{hyperref}

\def\aj{\rm{AJ}}                   
\def\apj{\rm{ApJ}}                 
\def\apjs{\rm{ApJS}}               
\def\aap{\rm{A\&A}}                
\def\mnras{\rm{MNRAS}}             
\def\prd{\rm{Phys.~Rev.~D}}        
\def\pasp{\rm{PASP}}               

\newcommand{\beq}{\begin{equation}}
\newcommand{\eeq}{\end{equation}}

\newcommand{\beqa}{\begin{eqnarray}}
\newcommand{\eeqa}{\end{eqnarray}}
\newcommand{\bea}{\begin{eqnarray}}
\newcommand{\eea}{\end{eqnarray}}

\title[DES Y1 Angular Power Spectrum]{Dark Energy Survey Year 1 Results:\\
Measurement of the Galaxy Angular Power Spectrum}

\author[DES Collaboration]{
\parbox{\textwidth}{
\Large
H.~Camacho$^{1,2}\star$,
N.~Kokron$^{1,2}\dagger$,
F.~Andrade-Oliveira$^{3,2}\ddagger$,
R.~Rosenfeld$^{4,2}$,
M.~Lima$^{1,2}$,
F.~Lacasa$^{5,2}$,
F.~Sobreira$^{6,2}$,
L.~N.~da Costa$^{2,7}$,
S.~Avila$^{8}$,
K. ~C.~Chan$^{9,10}$,
M.~Crocce$^{9,10}$,
A.~J.~Ross$^{11}$,
A.~Troja$^{4,2}$,
J.~Garc\'ia-Bellido$^{12}$,
T.~M.~C.~Abbott$^{13}$,
F.~B.~Abdalla$^{14,15}$,
S.~Allam$^{16}$,
J.~Annis$^{16}$,
R.~A.~Bernstein$^{17}$,
E.~Bertin$^{18,19}$,
S.~L.~Bridle$^{20}$,
D.~Brooks$^{14}$,
E.~Buckley-Geer$^{16}$,
D.~L.~Burke$^{21,22}$,
A.~Carnero~Rosell$^{2,7}$,
M.~Carrasco~Kind$^{23,24}$,
J.~Carretero$^{25}$,
F.~J.~Castander$^{9,10}$,
R.~Cawthon$^{26}$,
C.~E.~Cunha$^{21}$,
C.~B.~D'Andrea$^{27}$,
J.~De~Vicente$^{28}$,
S.~Desai$^{29}$,
H.~T.~Diehl$^{16}$,
P.~Doel$^{14}$,
J.~Estrada$^{16}$,
A.~E.~Evrard$^{30,31}$,
B.~Flaugher$^{16}$,
P.~Fosalba$^{9,10}$,
J.~Frieman$^{16,26}$,
D.~W.~Gerdes$^{30,31}$,
T.~Giannantonio$^{32,33,34}$,
R.~A.~Gruendl$^{23,24}$,
J.~Gschwend$^{2,7}$,
G.~Gutierrez$^{16}$,
D.~L.~Hollowood$^{35}$,
K.~Honscheid$^{11,36}$,
B.~Hoyle$^{37,34}$,
D.~J.~James$^{38}$,
M.~W.~G.~Johnson$^{24}$,
M.~D.~Johnson$^{24}$,
S.~Kent$^{16,26}$,
D.~Kirk$^{14}$,
E.~Krause$^{39,40}$,
K.~Kuehn$^{41}$,
N.~Kuropatkin$^{16}$,
H.~Lin$^{16}$,
J.~L.~Marshall$^{42}$,
R.~Miquel$^{43,25}$,
W.~J.~Percival$^{44,45}$,
A.~A.~Plazas$^{40}$,
A.~K.~Romer$^{46}$,
A.~Roodman$^{21,22}$,
E.~Sanchez$^{28}$,
M.~Schubnell$^{31}$,
I.~Sevilla-Noarbe$^{28}$,
M.~Smith$^{47}$,
R.~C.~Smith$^{13}$,
M.~Soares-Santos$^{48}$,
E.~Suchyta$^{49}$,
M.~E.~C.~Swanson$^{24}$,
G.~Tarle$^{31}$,
D.~Thomas$^{8}$,
D.~L.~Tucker$^{16}$,
A.~R.~Walker$^{13}$,
J.~Zuntz$^{50}$
\begin{center} (DES Collaboration) \end{center}
}
\vspace{0.4cm}
\\
\parbox{\textwidth}{
(affiliations are listed at the end of the paper)\\
$\star$ corresponding author: \href{mailto:hcamacho@if.usp.br}{hcamacho@if.usp.br}\\
$\dagger$ corresponding author: \href{mailto:nickolas.kokron@usp.br}{nickolas.kokron@usp.br}\\
$\ddagger$ corresponding author: \href{mailto:felipe.andrade@linea.gov.br}{felipe.andrade@linea.gov.br}
}
}

\begin{document}
\pagerange{\pageref{firstpage}--\pageref{lastpage}} \pubyear{0000}

\maketitle

\label{firstpage}

\begin{abstract}
We use data from the first-year (Y1) observations of the Dark Energy Survey (DES) collaboration to measure the galaxy angular power spectrum, and search for its baryonic acoustic oscillations (BAO) feature using a template-fitting method. We test our methodology in a sample of 1800 DES Y1-like mock catalogs. The angular power spectrum ($C_\ell$) is measured with the pseudo-$C_\ell$ method, using pixelized maps constructed from the mock catalogs and the DES mask. The covariance matrix of the $C_\ell$'s in these tests are also obtained from the mock catalogs. 
We use templates to model the measured spectra and estimate template parameters firstly from the $C_\ell$'s of the mocks using two different methods, a maximum likelihood estimator and a Markov Chain Monte Carlo method, finding consistent results with a good reduced $\chi^2$. Robustness tests are performed to estimate the impact of different choices of settings used in our analysis.  
After these tests on mocks, we apply our method to a galaxy sample constructed from DES Y1 data specifically for large scale structure studies. This catalog comprises galaxies within an effective area of 1318 
deg$^2$ and redshifts in the range $0.6<z<1.0$. 
We fit the observed spectra with our optimized templates, considering models with and without 
BAO features. 
We find that the DES Y1 data favors a model with BAO wiggles at the $2.6\,\sigma$ confidence level with a best-fit shift parameter of $\alpha=1.023\pm 0.047$. 
However, the goodness-of-fit is somewhat poor, with $\chi^2/$(dof) = 1.49. We identify a possible cause of this issue and show that using a theoretical covariance matrix obtained from $C_\ell$'s that are better adjusted to data results in an improved value of $\chi^2/$(dof) = 1.36 which is similar to the value obtained with the real-space analysis.   
Our results correspond to a measurement of the ratio of the angular diameter distance to the effective redshift of our sample, $z_{\rm eff} = 0.81$ and the BAO physical scale $r_d$ of $D_A(z_{\rm eff} = 0.81)/r_d = 10.65 \pm 0.49$, consistent with the main DES BAO findings.
This is a companion paper to the main DES BAO article showing the details of the harmonic space analysis.
\space
\end{abstract}


\space

\section{Introduction}\label{sec:intro}
\space

The large-scale distribution of galaxies carries information about the cosmological model that best describes our universe \citep[e.g.,][]{Dodelson:2003ft,Lyth:2009zz}.
After the great success of maps of the cosmic microwave background (CMB) in providing cosmological information, large galaxy surveys have become one of the major contributors to our understanding of gravity and the ingredients that make up the cosmos.
They provide evidence for the consistency of our description for the evolution of the universe from the early CMB epoch to present times.

The distribution of galaxies in the universe carries cosmological information that was imprinted from the era when baryons and photons were tightly coupled.
The so-called baryon acoustic oscillation (BAO) feature results from processes that occur up to the baryon drag epoch, and are sensitive in particular to the sound horizon $r_s$  at decoupling.

It is possible to quantify the distribution of galaxies by measuring its 2-point correlation function. 
One can measure the three-dimensional 2-point galaxy correlation function either in real space or measure its Fourier transform, the power spectrum, in harmonic space. In principle both quantities carry the same information, but in practice they may have different sensitivities to the estimation of cosmological parameters due to, among other effects, different covariance matrices, different response to systematic effects, etc.
For instance, gaussian covariance matrices for the power spectrum are diagonal in the full-sky case, whereas for the spatial correlation function significant correlations are expected.
Hence performing measurements in both real and Fourier space serves as a consistency check, and may also provide complementary information to tame some of the observational issues.

In galaxy surveys where redshifts are not precisely measured, as is the case with photometric redshifts (photo-$z$), one actually considers the projected galaxy distribution into redshift bins.
In this case what is measured is the angular correlation function in real space (ACF, denoted by $w(\theta)$) and/or the angular power spectrum in harmonic space (APS, denoted by $C_\ell$ in the following).

The APS is studied in the present work, which uses data from the first year (Y1) of observations from the Dark Energy Survey \citep[DES,][]{Flaugher:2004vg}, a large photometric survey in five bands that is planned to cover $5000$ deg$^2$ of the sky in a 5-year campaign.
The DES uses the Dark Energy Camera \citep[DECam,][]{Fla15}, a $570$-Megapixel camera mounted on the 4-meter Blanco telescope at the Cerro Tololo Inter-American Observatory, Chile and is currently in its fifth year of data acquisition.
The DECam received its first light in September 2012, followed by a Science Verification (SV) period covering an area of approximately 250 deg$^2$. 
Measurements of the ACF and the impact of systematic errors in the SV data were reported in \cite{Cro16}.
More recently, cosmological results from combined clustering and weak lensing measurements in the DES Y1 data have been presented \citep{Abbott:2017wcz}.

The BAO feature in the 2-point galaxy correlation function has been observed in several surveys. A few examples are the 2-degree Field Galaxy Redshift Survey (2dFGRS) \citep{Percival:2001hw,Cole:2005sx}, the Sloan Digital Sky Survey (SDSS) I, II, III and IV \citep{Eis05,Padmanabhan:2012hf, Anderson:2013oza,Ross:2014qpa,Alam:2016hwk,Ata:2017dya} and the WiggleZ survey \citep{Blake:2011en}.
In particular, the BAO scale was measured in SDSS photometric samples using the ACF \citep{Carnero:2011pu,2011MNRAS.416.3033S,deSimoni:2013oqa} and the APS \citep{Blake:2006kv,Seo:2012xy}.

In this work we use a template-based method to study the BAO feature in the angular power spectra from the DES Y1 data. We describe our method and test it on realistic survey mocks. These mocks were also used to measure the covariance matrix of the $C_\ell$'s.
The covariance matrix was then used to find the likelihood corresponding to the template adopted to model the data. We estimate the significance of the detection of the BAO feature for a baseline template using two independent methods: a maximum likelihood estimator (MLE) and a Markov Chain Monte Carlo (MCMC) method. We also present the reduced $\chi^2$ values for the mocks to demonstrate the goodness-of-fit.
We explore the robustness of our baseline model to the estimation of parameters testing different choices of settings and assumptions in the analysis.
After the validation of our methodology we apply it to Y1 data with the intent to search for BAO features.
We find that 
the DES Y1 data favors a model with BAO wiggles at the $2.6\,\sigma$ confidence level with a best-fit shift parameter of $\alpha=1.023\pm 0.047$ with a somewhat large value of $\chi^2$/(dof) = 1.49. We investigate this issue substituting the covariance matrix obtained from the mocks by a gaussian theoretical covariance matrix taking into account the Y1 mask with input $C_\ell$'s that are better adjusted to data obtaining an improved value of $\chi^2/$(dof) = 1.36 which is similar to the value obtained with the real-space analysis. 

This paper is part of a series related to the detection of the BAO features in Y1 data. It relies on the construction of a catalog suitable for the study of clustering of galaxies, especially concerning the BAO feature \citep{Crocce:2017iwq}, the mock catalogs used to validate the analysis and results \citep{Avila:2017nyy}
and the computation of galaxy photo-$z$s (Gaztanaga et al. in prep). Other papers detail methods to study the BAO feature in configuration space with the angular correlation function $w(\theta)$ \citep{2018arXiv180104390C},
and using the comoving transverse separation \citep{Ross:2017emc},
while the present work details the use of the angular power spectrum. 
The joint results applied to the Y1 data are described in the BAO main paper \citep{Abbott:2017wau}. 

This paper is organized as follows.
We start by describing the theoretical modelling of the angular power spectrum in \S~\ref{sec:theory}, including the template used to study the BAO feature.
In \S~\ref{sec:catalog} we describe the DES Y1 galaxy catalog constructed for BAO studies, focusing on the redshift binning, pixelization and masking. The 1800 mock catalogs used for the verification of our measurements, for the covariance matrix estimation and for testing our parameter estimation from the template method are briefly presented in \S~\ref{sec:simulations}. 
In \S~\ref{sec:measurements} the measurement of the APS using the pixelized maps is described.
The methodology we adopt is tested on the mocks in \S~\ref{sec:methodology} where we also study the impact of different choices of templates and settings on the parameter estimation as robustness checks.
Having validated our methodology, we apply it for Y1 data in \S~\ref{sec:BAOY1} where we concentrate on finding BAO features in the angular power spectrum.
Finally, in \S~\ref{sec:conclusions} we present our conclusions. 

\section{Theory}\label{sec:theory}

In this Section we review basic concepts used for the theoretical modelling throughout the paper.

\subsection{Angular Power Spectrum and Theoretical Covariance Matrix}

We define the 3-dimensional matter overdensity $\delta_{\rm m}({\bf x})$ as
\bea
\delta_{\rm m}({\bf x})=\frac{\rho_{\rm m}({\bf x})-\bar{\rho}_{\rm m}}{\bar{\rho}_{\rm m}}\,,
\eea
where $\rho_{\rm m}({\bf x})$ is the matter density at point ${\bf x}$ and $\bar{\rho}_{\rm m}$ is the background matter density. 

We decompose ${\bf x}=\chi(z)\hat{\bf n}$ where $\hat{\bf n}$ is the angular direction and $\chi(z)$ is the comoving angular-diameter distance at redshift $z$.
Since we only consider flat-universe cosmologies, $\chi(z)$ is simply the comoving radial distance to redshift $z$. 

In Fourier space, the overdensity is 
\bea
\delta_{\rm m}({\bf k}) =\int d^3{x} \ e^{-i{\bf k}\cdot {\bf x}} \delta_{\rm m}({\bf x})\,,
\eea
and it defines the 3-dimensional matter power spectrum $P_{\rm m}(k)$ by the relation
\bea
\langle \delta_{\rm m}({\bf k})  \delta_{\rm m}^*({\bf k}^\prime) \rangle = (2\pi)^3 \delta^3({\bf k}-{\bf k}^\prime) P_{\rm m}(k)\,.
\eea

On linear scales, we assume that the 3-dimensional {\it galaxy} overdensity $\delta_{\rm gal}({\bf x})$ is related to the matter overdensity by
\bea
\delta_{\rm gal}({\bf x}) = \frac{n_{\rm gal}({\bf x})-\bar{n}_{\rm gal}}{\bar{n}_{\rm gal}} = b(z) \delta_{\rm m}({\bf x})\,,
\eea
where $n_{\rm gal}({\bf x})$ is the galaxy number density at ${\bf x}$, $\bar{n}_{\rm gal}$ is its mean value and $b$ is the scale-independent linear galaxy bias. 
Therefore one has the relation
$P_{\rm gal}(k)=b^2 P_{\rm m}(k)$. 

For a normalized galaxy selection function $\phi^i(z)=dN/dz$ at photo-z bin $i$, we define the projected 2-dimensional galaxy overdensity  $\delta_{{\rm gal}}^i(\hat{\bf n})$ as
\bea
\delta_{{\rm gal}}^i(\hat{\bf n}) =\int dz \ \phi^i(z)   \delta_{{\rm gal}}^i({\bf x})\,.
\eea

This galaxy overdensity can be decomposed into spherical harmonics $Y_{\ell m}$ as
\bea
\delta_{{\rm gal}}^i (\hat{\bf n}) = \sum_{\ell=0}^\infty \sum_{m=-\ell}^{\ell} a_{\ell m}^i Y_{\ell m}(\hat{\bf n})\,,
\label{eq:decomp}
\eea
where $a_{\ell m}$ are the harmonic coefficients. The angular cross-spectrum $C_{\ell}^{ij}$ is defined via
\bea
\langle (a_{\ell m}^i) (a_{\ell' m'}^{j \hspace{0.0in}})^* \rangle \equiv \delta_{\ell \ell'}\delta_{mm'} C_{\ell}^{ij}\,,
\label{eq:Cl}
\eea
and the angular power spectrum at $z$ bin $i$ is defined by $C_{\ell}^i\equiv C_{\ell}^{ii}$. 
One can write the spherical harmonics coefficients as \citep[see e.g.][]{2011MNRAS.414..329C, Sobreira:2011an}
\bea
a_{\ell m}^i = \int dz  \ \phi^i(z) \int \frac{d^3k}{(2 \pi)^3} \ \delta_{\rm gal}({\bf k},z)   4 \pi  i^\ell  j_\ell(k\chi)   Y^\ast_{\ell m}(\hat{\bf k}) \,, 
\label{eq:alm}
\eea
where in linear theory $\delta_{\rm gal}({\bf k},z) = G(z)\delta_{\rm gal}({\bf k}, 0)$ and $G(z)$ is the linear growth function normalised such that $G(0)=1$. 
Hence, from Eqs.~(\ref{eq:Cl}) and (\ref{eq:alm}) we can write $C_{\ell}^i$ as:
\bea
C_{\ell}^i = \frac{2}{\pi} \int dk \ k^2 P_{\rm gal}(k) \left\{ \Psi^i_\ell \right\}^2\, ,
\label{eq:Cell_theory}
\eea
where
\bea
\Psi^i_\ell = \int dz \ G(z)\phi^i(z)j_\ell[k\chi(z)]\,.
\label{eq:Psi}
\eea

A similar symmetrized expression holds for $C_\ell^{ij}$, replacing $\left\{ \Psi^i_\ell \right\}^2$ by $\Psi^i_\ell \Psi^j_\ell$ in Eq. (\ref{eq:Cell_theory}).

We include linear redshift space distortions in our fiducial modeling by modifying $\Psi^i_\ell $ to :
\bea
&&\Psi^i_\ell = \int  dz \ \beta(z) G(z)\phi^i(z) 
\left\{ \frac{2 \ell^2 + 2 \ell - 1}{(2 \ell +3)(2 \ell -1)}j_\ell[k\chi] - 
 \right. \nonumber \\
&& \left.  \frac{\ell ( \ell - 1)j_{\ell-2} [k\chi]}{(2 \ell -1)(2 \ell +1)} - 
\frac{(\ell+1) ( \ell +2)}{(2 \ell +1)(2 \ell +3)}j_{\ell+2} [k\chi] \right\}\,,
\label{eq:Psi_RSD}
\eea
where
\bea
\beta(z) = \frac{1}{b(z)} \frac{d \ln G}{d \ln a}.
\eea

The Gaussian covariance matrix for $C_\ell$'s measured at photo-$z$ bins $i$ and $j$ can be theoretically modeled in the so-called $f_{\rm sky}$ approximation as:
\bea
{\rm Cov}[{\rm APS}]_{\ell \ell^\prime}^{ij} &\equiv& \langle C_{\ell}^i C_{\ell^\prime}^j \rangle 
- \langle C_{\ell}^i \rangle\langle C_{\ell^\prime}^j \rangle \nonumber \\
&=& \frac{2 }{f_{\rm sky} \Delta \ell (2\ell+1)}   \left( C_{\ell}^{ij} + \frac{\delta_{ij}}{\bar{n}_i} \right)^2 \delta_{\ell\ell^\prime} \,, \, \hspace{0.2in} 
\label{eq:cov}
\eea
where $\Delta \ell$ is the $\ell$ bin size,  $f_{\rm sky}$ 
is the sky fraction covered by the survey, $\bar{n}_i$ is the mean galaxy number density at bin $i$ and $\delta_{ij}$ is a Kronecker delta.  

The analytical expression we actually use to estimate the theoretical covariance is more realistic, as it is tied to the pseudo-$C_{\ell}$ estimator (see \S~\ref{sec:measurements}) and corrects for binning and mask effects \citep{Efstathiou:2003dj}. Interestingly, we find that the above approximation multiplied by a boost factor agrees well with the full covariance expression and with the covariance estimated from mock catalogs in the range of $\ell$ explored in this work (see \S~\ref{subsec:cov}).

\subsection{$C_\ell$ Template}\label{subsec:templates}

Our goal is to extract from mocks and from DES Y1 observations the scale associated with the BAO feature, namely the angular distance scale $D_A(z)$. In order to be as insensitive as possible to nonlinear effects such as bias and redshift space distortions, we will use a template method \citep{Seo:2012xy,2014MNRAS.441...24A,2016MNRAS.460.4210G,Ata:2017dya,Ross:2017emc,2018arXiv180104390C}.

Since the selection function for the simulations and data is fully specified, the $C_\ell$ template may be defined by first settling on a template for $P(k)$ and projecting onto $C_\ell$'s using Eq.~(\ref{eq:Cell_theory}). 
We start with 
\begin{equation}
P^{\rm temp}(k) = [ P(k)^{\rm lin} - P^{\rm nw}(k)] e^{-k^2 \Sigma_{nl}^2} + P^{\rm nw}(k),
\end{equation}
where $P(k)^{\rm lin}$ is the linear power spectrum and the no-wiggle power spectrum $P(k)^{\rm nw}$ is obtained from the Eisenstein-Hu parametrization \citep{Eisenstein:1997ik}.
The nonlinear damping scale is fixed at $\Sigma_{nl} = 5.2$ Mpc/h, which was determined from a fit to the mean of mocks \citep{2018arXiv180104390C}. 

We chose our template by optimizing the BAO signal in the mock catalogs. Tests of different templates will be shown below.
Our default template for $C_\ell$ is given by:
\bea
C^{\rm }(\ell) &=& B_0 \, C^{\rm temp}(\ell/\alpha) +   A_{0} + A_{1} \ell + A_{2}/\ell^2 \,,
\eea
where $C^{\rm temp}(\ell)$ is the projection of $P^{\rm temp}(k)$ as described above. The amplitude $B_0$ is related to the linear bias squared and the parameters $A_i$ take into account scale-dependent bias, shot noise, uncertainties in the redshift-space distortions, etc. We allow these parameters to change with redshift.
Therefore, for 4 redshift bins we will have 16 of these parameters to adjust.
We will marginalize over them in MCMC analysis and keep them fixed at the values that maximize the likelihood in the MLE analysis as will be described in section \ref{sec:methodology}.

The most important parameter in our analysis is the so-called shift parameter $\alpha$, which measures the shift of the BAO peak positions with respect to a fiducial cosmology. We will assume that it does not change significantly with respect to its value at the ``effective redshift" of the sample used ($z_{\rm eff} = 0.81$ in the BAO Y1 sample) \citep{Abbott:2017wcz}. 
If the fiducial cosmology used to compute $P(k)^{\rm lin}$ and $P(k)^{\rm nw}$ turns out to be the correct one
then one should find $\alpha=1$. The shift parameter is related to the change in the BAO location, given by the ratio of the angular diameter distance $D_A(z)$ to the sound horizon scale at the drag epoch ($r_d$):
\bea
\alpha = \frac{\left(D_A(z)/r_d\right)}{\left(D_A(z)/r_d\right)_{\rm fid}}.
\eea
For example, for a fiducial cosmology given by the MICE simulations ($h=0.7,\ \Omega_m=0.25,\ \Omega_\Lambda = 0.75$)  \citep{Cro10, Fosalba:2013wxa} we find
that, with respect to the cosmology found by DES combined with other observations ($h= 0.678 ,\ \Omega_m=0.30, \ \Omega_\Lambda = 0.70$) \citep{Abbott:2017wau}, $\alpha \approx 1.03$.

We will test this parametrization with the mocks below and show that it results in biases below $1 \%$ for the parameter estimation. We study the impact of other templates as robustness tests in \S~\ref{sec:methodology}.

\section{DES Y1 Galaxy Catalog}\label{sec:catalog}

\subsection{Catalog}

The  catalog for LSS analyses using DES Y1 data was created from the so-called Y1 Gold catalog \citep{Drlica-Wagner:2017tkk}
which in turn was built from the data reduction performed by the Dark Energy Survey Data Management (DESDM) system on DECam images. The LSS sample selection is based on color, magnitude and redshift cuts designed to provide an optimal balance between the density of objects and the photometric redshift uncertainty for $z > 0.6$, minimizing the forecasted BAO error \citep{Crocce:2017iwq}.  
We will use the LSS catalog with photometric redshifts obtained with a Multi-Object Fitting (MOF) photometry \citep{Drlica-Wagner:2017tkk} and the Directional Neighborhood Fitting (DNF) algorithm \citep{DeVicente:2015kyp}. After proper masking described in \citep{Crocce:2017iwq} the catalog has approximately 1.3 million galaxies in an area of $1317$ deg$^2$.

\begin{figure}
\hspace*{-0.3cm}
\centering
\includegraphics[width=\linewidth]{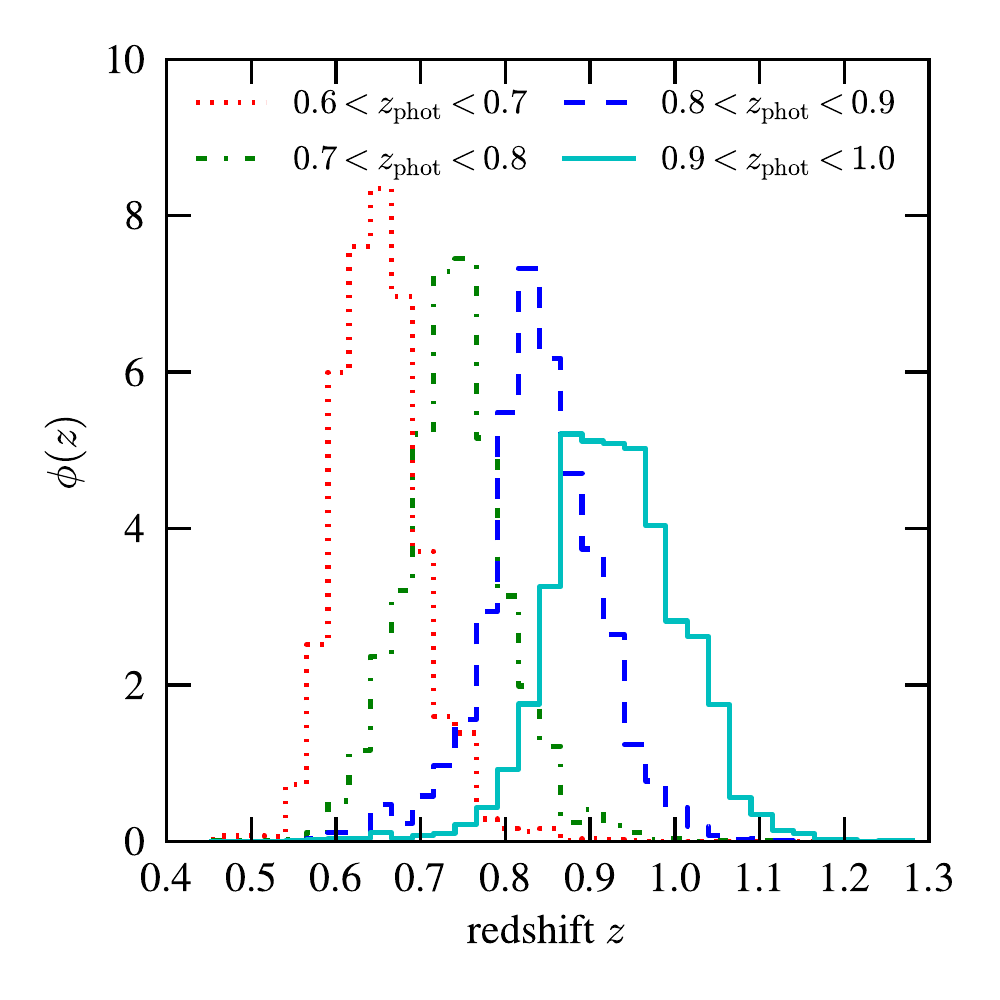}
\caption{Redshift selection function $\phi(z) \propto dN/dz$ in the four photo-$z$ bins considered 
in this work. 
}
\label{fig:dndz}
\end{figure}

We divide the catalog into 4 tomographic photo-$z$ bins with width $\Delta z_{\rm phot} = 0.1$ in the range $0.6 < z_{\rm phot} < 1.0$. In Fig.~\ref{fig:dndz} we show the redshift distribution for each bin obtained by stacking a Monte Carlo sampled value of 
the photo-$z$ from the estimated probability distribution function for each object.
The bins are defined using a point-estimate of the photo-$z$ given by the maximum likelihood redshift produced by DNF.

The LSS catalog also comes with a correction for the three main systematic dependencies found on observational quantities: local stellar density, mean $i$-band PSF (FWHM) and detection limit ($g$-band depth). These corrections are encapsulated in a weight factor for each object, which is applied to reduce these dependencies. 

\subsection{Pixelized map generation}
Each redshift bin is pixelized using HEALPix \citep{Gorski:2004by} at a resolution of $N_{\rm side} = 1024$, with $N_I$ galaxies in each pixel $I$.
A pixelized angular mask described in  \citep{Crocce:2017iwq}
is used to find the density contrast maps. 
Each pixel in the mask comes with its fractional coverage which we denote $w_I$ such that
\begin{equation}
\frac{\sum_I w_I}{N_{\rm pix}} = f_{\rm sky}\,,
\end{equation}
where $N_{\rm pix} = 12 N_{\rm side}^2$ is the total number of pixels for a given resolution and again $f_{\rm sky}$ is the survey sky fraction.

We degrade the mask resolution from  $N_{\rm side} = 4096$ to $N_{\rm side} = 1024$ keeping the fractional coverage as an average of the smaller pixels contained in the large (smaller resolution) pixel.
The number density of galaxies $n_I$ in each pixel inside the footprint is computed as: 
\begin{equation}
n_I = \frac{N_I}{w_I \Omega}\,,
\end{equation}
where $\Omega$ is the (common) area of one pixel. 
The average number density of galaxies is
\begin{equation}
\bar{n} = \frac{\sum_I N_I}{(\sum_I w_I) \Omega}\,,
\end{equation}
and finally the density contrast $\delta_I$ in each pixel is given by:
\begin{equation}
\delta_I = \frac{n_I}{\bar{n}} - 1 \,.
\end{equation}

These maps generated for each redshift bin are used to measure the APS as explained in \S~\ref{sec:measurements}.

\section{DES Mock Simulations}\label{sec:simulations}

In addition to the DES Y1 galaxy catalog, we use a set of 1800 galaxy mock simulations, especially made for studies of large-scale structure in DES, including the present BAO analysis \citep{Avila:2017nyy}.

These mocks serve a dual purpose in our study. 
First, we use them to test our codes for estimating $C_\ell$'s, covariances and the BAO feature extraction in a DES-like survey. 
Second, we make direct use of the covariance matrices estimated from them in the BAO analysis of the DES Y1 data.

These simulations match all aspects of the DES Y1 data, including its footprint, abundance and clustering of galaxies and redshift distribution.
One starts with halo catalogs that are constructed with the HALOGEN method \citep{Avila:2014nia}, such that they satisfy halo mass-functions and bias appropriately from N-body simulations. Next, galaxies are assigned to these halos according to a hybrid Halo Occupation Distribution (HOD)/Halo Abundance Matching (HAM) prescription. 
The methodology is much faster than using full N-body simulations and allows for the construction of thousands of simulations. 
These mock catalogs were constructed using the MICE Grand Challenge N-body simulations \citep{Cro10, Fosalba:2013wxa,Crocce:2013vda,Fosalba:2013mra}, 
with cosmological parameters close but not equal to those of the Planck mission.
We refer the reader to \cite{Avila:2017nyy} for details of the construction of these DES galaxy mocks.

\section{Angular Power Spectrum Measurement in Cut Sky}\label{sec:measurements}

For data collected over the whole sky, an unbiased estimator of the APS is simply the average of the $a_{\ell m}$ coefficients over all $m$ values \citep{Hivon:2001jp}:
\begin{equation}
\hat{C}_\ell = \frac{1}{2\ell+1} \sum_{m=-\ell}^{m=\ell} |a_{\ell m}|^2\,.
\end{equation}

When performing full-sky estimations, we compute the coefficients $a_{\ell m}$ from the pixelized density contrast maps using the {\tt anafast} routine within HEALPiX.

As the DES measurements are not made over the full-sky, the previous estimator is not appropriate, since spherical harmonics no longer provide a complete orthonormal basis to expand angular overdensities.
In this case, we use the so-called pseudo-$C_\ell$ method to relate the APS measured in a  masked sky $\hat{C_\ell}$ to the ``true" APS $C_\ell$ \citep{Hivon:2001jp}. The pseudo-$C_\ell$ estimator relies on the assumption:
\begin{equation}
\langle \hat{C}_\ell (\tilde{\delta}_{\rm gal}(\hat{\bf n}) \rangle = \sum_{\ell'} {\cal M}_{\ell \ell'} C_{\ell'}(\delta_{\rm gal}(\hat{\bf n})),
\label{eq:pseudocl}
\end{equation}
where ${\cal M}$ is called the coupling matrix.
In the equation above the masked density contrast field $\tilde{\delta}_{\rm gal}$ is related to the full-sky one $\delta_{\rm gal}$ by a mask function $M$:
\begin{equation}
\tilde{\delta}_{\rm gal}(\hat{\bf n}) = M(\hat{\bf n}) \delta_{\rm gal}(\hat{\bf n})\,.
\end{equation}
It can be shown that the coupling matrix in terms of Wigner $3-j$ symbols is given by:
\begin{equation}
{\cal M}_{\ell_1\ell_2} =  (2 \ell_2 + 1) \sum_{\ell_3} \frac{2\ell_3+1}{4 \pi} C_{\ell_3}(M) 
\left( \begin{array}{ccc}
\ell_1 & \ell_2 & \ell_3 \\ 
0 & 0 & 0
\end{array} \right)^2\,,
\end{equation}
where  $ C_{\ell}(M)$ is the angular power spectrum of the pixelized mask.
Notice that for full sky measurements, ${\cal M}_{\ell \ell'}$ is simply the identity matrix and in general it carries information about the survey geometry and mask. 
The true $C_\ell$ can then be estimated from the pseudo-$C_\ell$ by solving the linear system defined by  Eq.~(\ref{eq:pseudocl}).

We use two independent codes to measure $C_\ell$'s via the pseudo-$C_\ell$ method without shot-noise subtraction. 
The first code is our own implementation of the pseudo-$C_\ell$ method in python. 
The second is the publicly available code {\tt NaMaster}\footnote{https://github.com/damonge/NaMaster}, which is implemented in C. 
We compared the $C_\ell$'s estimated from the two codes when applied to a single DES mock simulation.
The two codes agree at better than 5\% for all $\ell$ values considered here, and better than 1\% for $\ell > 100$, indicating our measurements are robust. All results presented in the remainder of the paper will make use only of the {\tt NaMaster} code. 

We consider in our default analysis multipoles in the range $30 < \ell <330$, corresponding roughly to the angular scales used in the $w(\theta)$ analysis \citep{2018arXiv180104390C},
and we then bin using a bin width of $\Delta \ell = 15$ in order to make the reduced covariance
matrix more diagonal and amenable to algebraic manipulations. Effects of different ranges and binnings will be studied as robustness tests in \S~\ref{subsec:rob}.

Finally, the covariance matrix is estimated from the $N_{\rm m}=1800$ mocks as:
\bea
{\rm Cov}[{\rm APS}]_{\ell \ell^\prime}^{ij}
= \frac{1}{N_{\rm m} - 1} \sum_{k=1}^{N_{\rm m}} \left( C_{\ell}^{i \hspace{0.01in} (k)} - \bar{C}_{\ell}^{i} \right)  
\left( C_{\ell^\prime}^{j \hspace{0.01in} (k)} - \bar{C}_{\ell^\prime}^j \right) \,,
\eea 
where the average $\bar{C}_{\ell}^i$ at photo-$z$ bin $i$ is given by
\bea
\bar{C}_{\ell}^i = \frac{1}{N_{\rm m}} \sum_{k=1}^{N_{\rm m}} C_{\ell}^{i \hspace{0.01in} (k)} .
\eea

\section{Tests of methodology on mocks}\label{sec:methodology}

We now apply our full methodology on the 1800 DES Y1 HALOGEN mock simulations with known cosmology and perform robustness tests to estimate the impact of changing our default settings on parameter estimation.

\subsection{Measurements of the APS}
\begin{figure}
\hspace*{-0.3cm}
\centering
    \includegraphics[width=\linewidth]{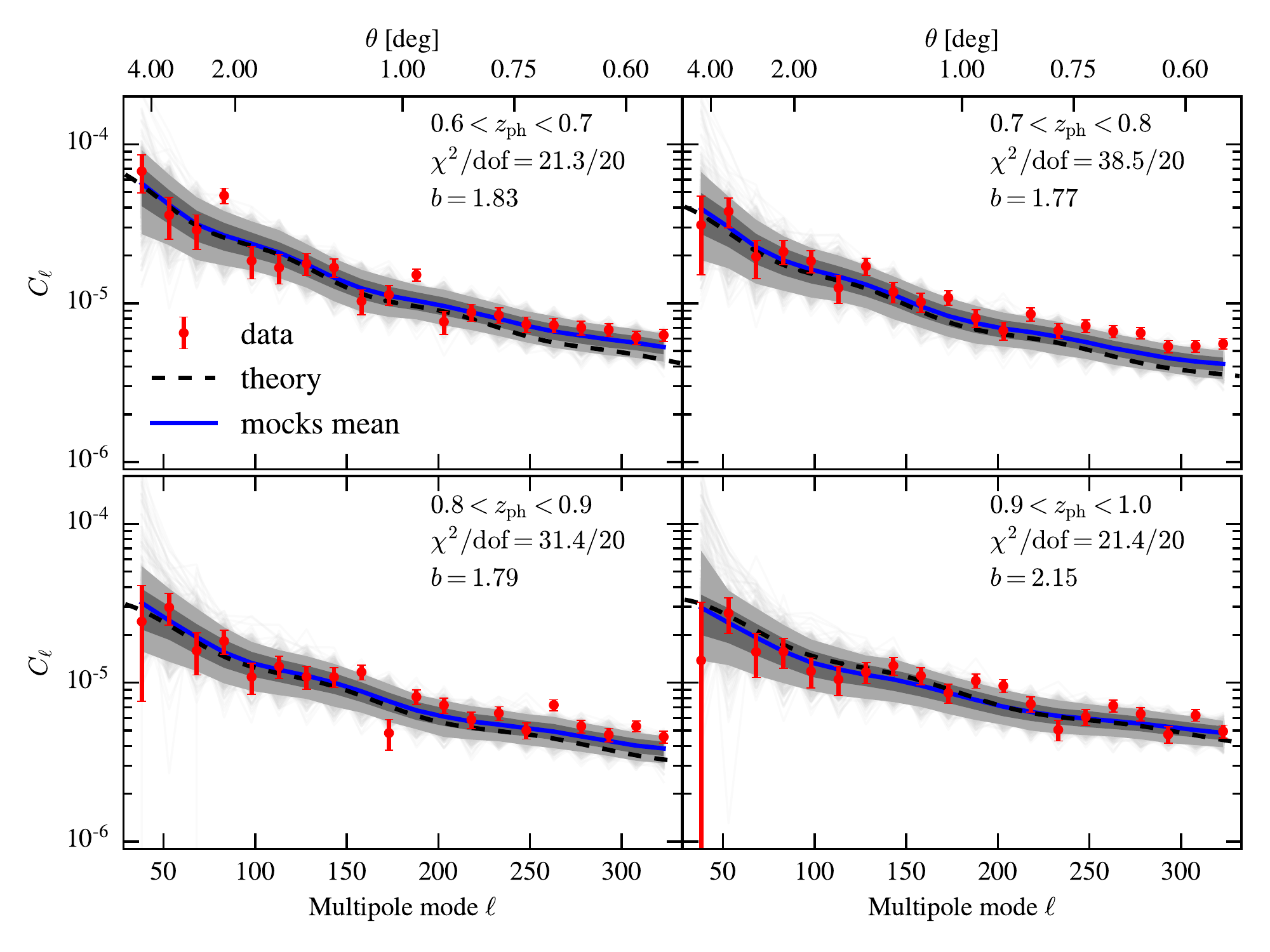}
    \caption{Measurements of $C_\ell$ in four photo-$z$ bins for the 1800 mocks (gray lines) and the Y1 data (red circles).
      Dashed line shows the theoretical prediction from a linear spectrum with MICE cosmology multiplied by a bias factor (shown in the panels) and including shot-noise and shaded regions show $68\%$ and $95\%$ C.L. from mocks measurements.  The blue line is the average of the mocks. The $\chi^2$ values show reasonable agreement between average measurements of the mocks and measurements on data.} 
  \label{fig:MocksDataTheory}
\end{figure}

In Fig.~\ref{fig:MocksDataTheory} we show the results of our $C_\ell$ measurements in the four photo-$z$ bins for the 1800 mocks together with the mean of the mocks. 
We also show theoretical predictions from $C_\ell$'s computed with a linear matter spectrum at the same cosmology as the mocks. In each photo-$z$ bin, we multiply the theoretical matter $C_\ell$'s by a galaxy bias factor squared determined by \citep{Avila:2017nyy} and add a shot-noise determined by the number density of Y1 galaxies in that photo-$z$ bin. 

The measured $C_\ell$'s from the mocks are in good agreement with the theoretical prediction. However, when compared to data there is some discrepancy in the second and third redshift bins, reflected in the somewhat high value of $\chi^2 = 1.92$ and $1.57$ respectively. In these bins the $C_\ell$'s from data exceed the ones from the mocks at large $\ell$'s. We will discuss some consequences of this behaviour below.

\begin{figure*}
\hspace*{-0.3cm}
\centering
    \includegraphics[width=0.5\linewidth]{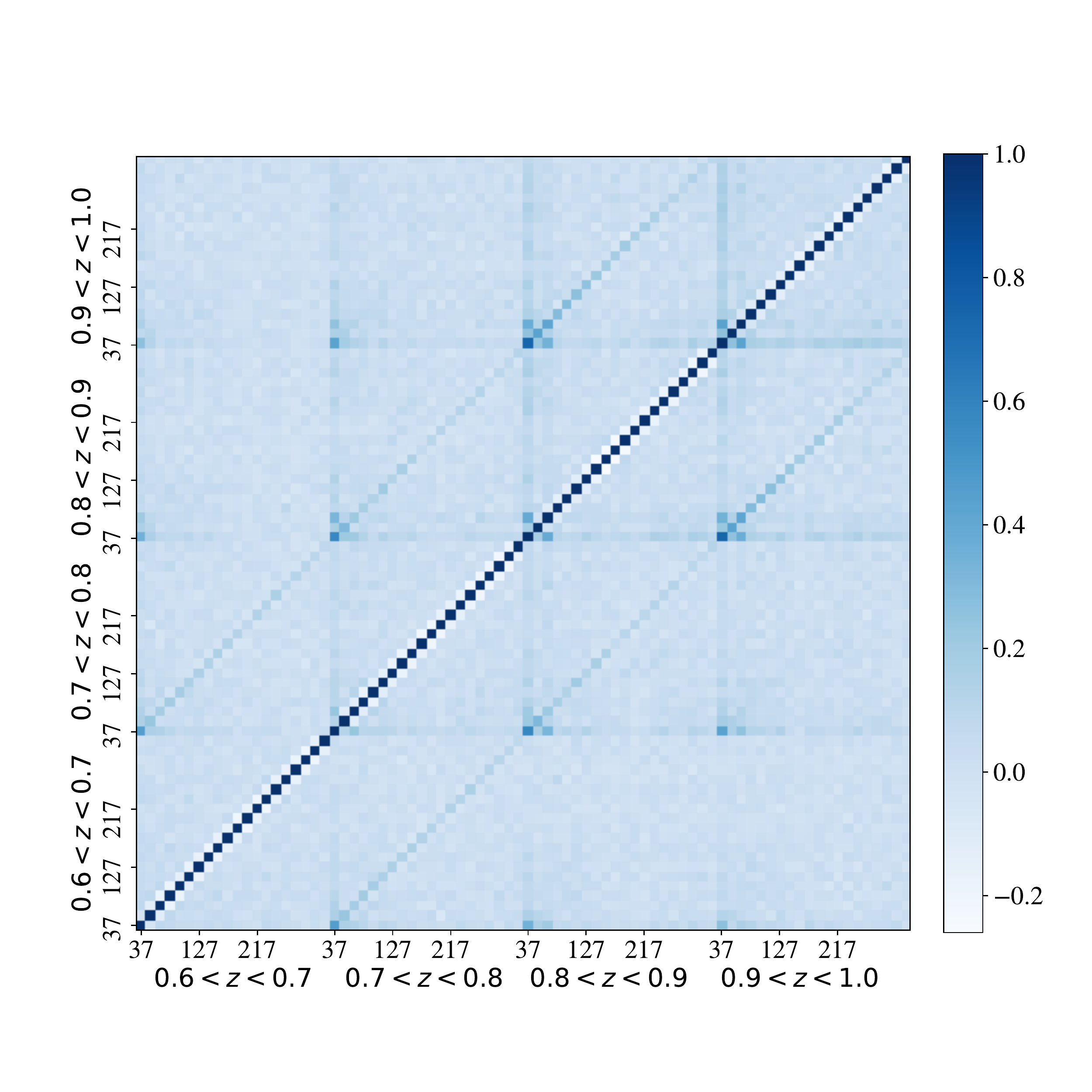}
    \includegraphics[width=0.5\linewidth]{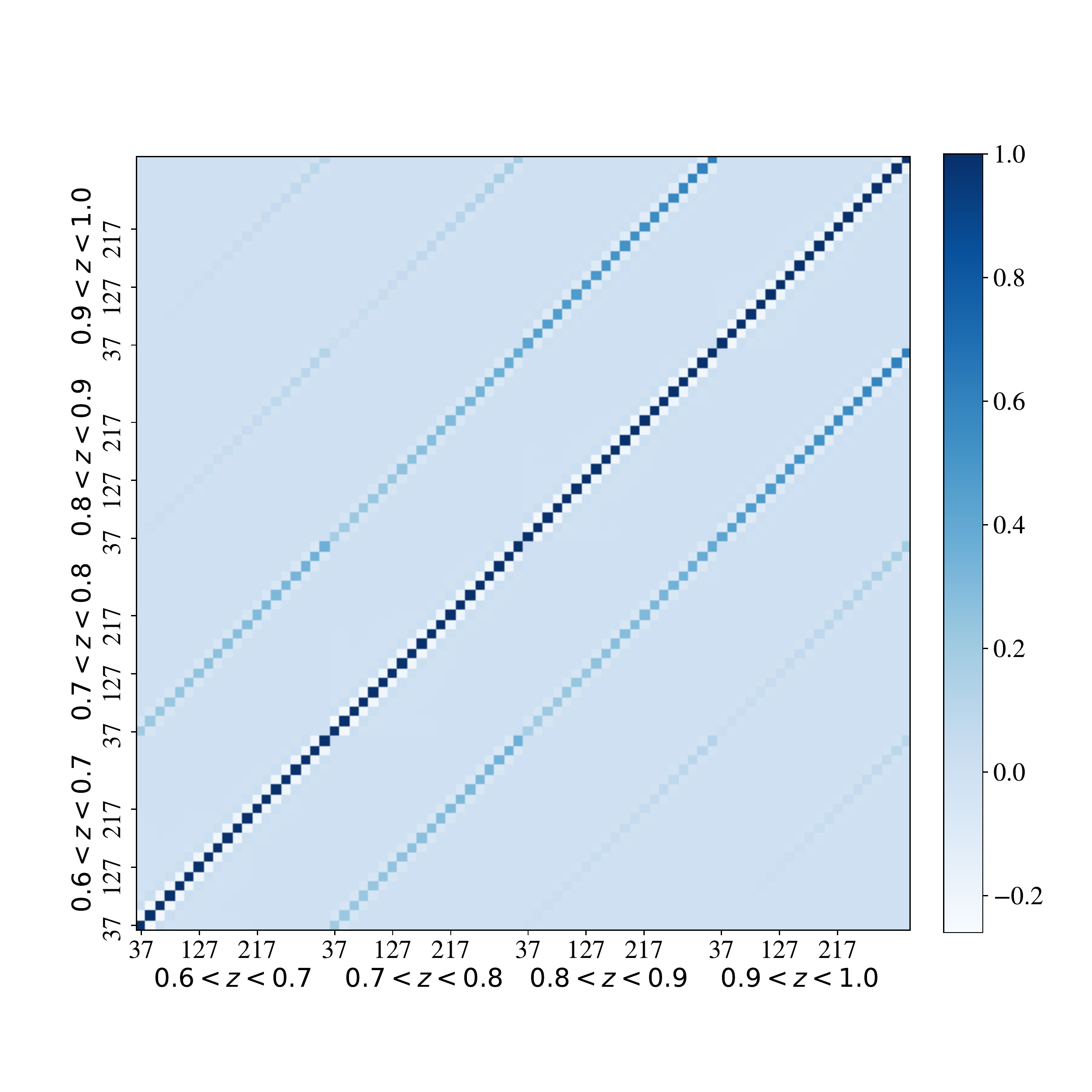}
  \caption{$C_\ell$'s correlation matrix for the 4 photo-$z$ bins. {\it Left}: measured from the 1800 mock simulations mimicking the DES Y1 data. 
  {\it Right}: Theoretical estimation computed at the mock cosmology, and accounting for binning and mask effects.} 
  \label{fig:Covzbins}
\end{figure*}  

\subsection{Covariance matrix}
\label{subsec:cov}

In order to quantify the correlation between bandpowers in our analysis, 
we show the correlation matrix,
\begin{equation}
\label{eq:corrmat}
{\rm Corr}_{ab} = \frac{{\rm Cov}_{ab}}{\sqrt{{\rm Cov}_{ab} \times{\rm Cov}_{ab}}},
\end{equation}
where the $a,b$ indexes label bandpowers, as measured from the mocks in the left panel of Fig.~\ref{fig:Covzbins} for the four redshift bins and using our fiducial $\Delta \ell = 15$ binning.
We see it is close to block-diagonal with structure similar to the one found in \citet{2018arXiv180104390C} 
for the covariance matrix for $w(\theta)$.  
The right panel of Fig.~\ref{fig:Covzbins} shows the theoretical estimation for the covariance computed at the mock cosmology, using the mock bias, adding the data shot-noise and correcting for binning and mask effects \citep{Efstathiou:2003dj}. The theoretical covariance is much less noisy, as expected.

\begin{figure}
\hspace*{-0.3cm}
\centering
    \includegraphics[width=\linewidth]{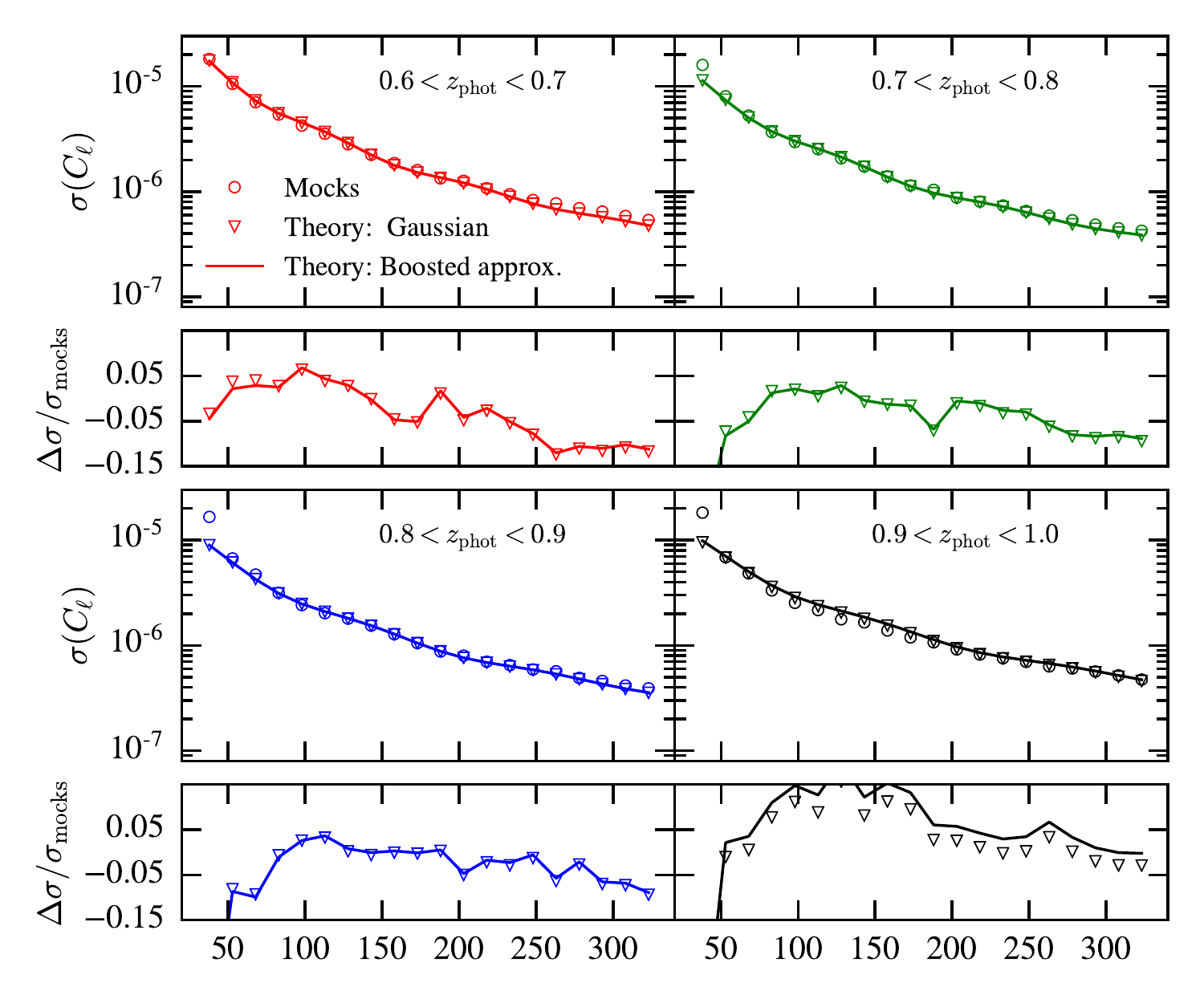}
    \caption{Comparison of the $C_\ell$'s diagonal errors in 4 photo-$z$ bins.
	For each bin, the top panel shows the standard deviation estimated from the 1800 DES mock simulations (open circles) the Gaussian prediction on the fiducial cosmology (solid lines) after rescaled by an empirical boost factor of $1.35$ and the Gaussian prediction from the pseudo-$C_\ell$ method (open triangles).
  The bottom panels show the relative differences with respect to the mocks standard deviation.} 
  \label{fig:Cov}
\end{figure}

In Fig.~\ref{fig:Cov}, we compare the diagonal errors of the $C_\ell$'s estimated from the mock simulations and the Gaussian prediction of the fiducial cosmology using two approximations: the naive $f_{\rm sky}$ approximation Eq.~(\ref{eq:cov}) and the prediction of the covariance matrix of the pseudo-$C_\ell$ estimator \citep{Efstathiou:2003dj,Brown:2004jn}.
We find good agreement between the errors coming from the simulation covariance matrix and from the pseudo-$C_\ell$ estimator.
However, for  the $f_{\rm sky}$ approximation we find that a ``boost factor" of $1.35$ is necessary to match the measured errors.
This was also the case for a similar analysis in SDSS \citep{2012ApJ...761...14H}.

We will use the full covariance matrix estimated from the mocks.
It is well known that statistical noise on the estimation of the covariance matrix from mock realizations translates into a bias on its inverse, the precision matrix, which is the actual fundamental piece on the likelihood estimation. We included this correction factor in our analysis \citep{Hartlap07, Dodelson2013, Percival2014}.
Given the number of mocks used, we have checked that the correction factor for the precision matrix is always less than 5\%, having no impact on the recovered value of $\alpha$.

\subsection{Parameter estimation}

We use two independent methods for the parameter estimation: a Markov Chain Monte Carlo (MCMC) implemented with {\tt emcee} \citep{2013PASP..125..306F} and a maximum likelihood estimator (MLE) with analytical least square fit of the nuisance parameters \citep{Cowan:1998ji}. We used our default BAO template described in \S~\ref{subsec:templates} with the covariance matrix estimated from the mocks. We performed a joint fit for the 4 photo-$z$ bins with 17 parameters in our default template.

\begin{figure}
\hspace*{-0.3cm}
\centering
    \includegraphics[width=\linewidth]{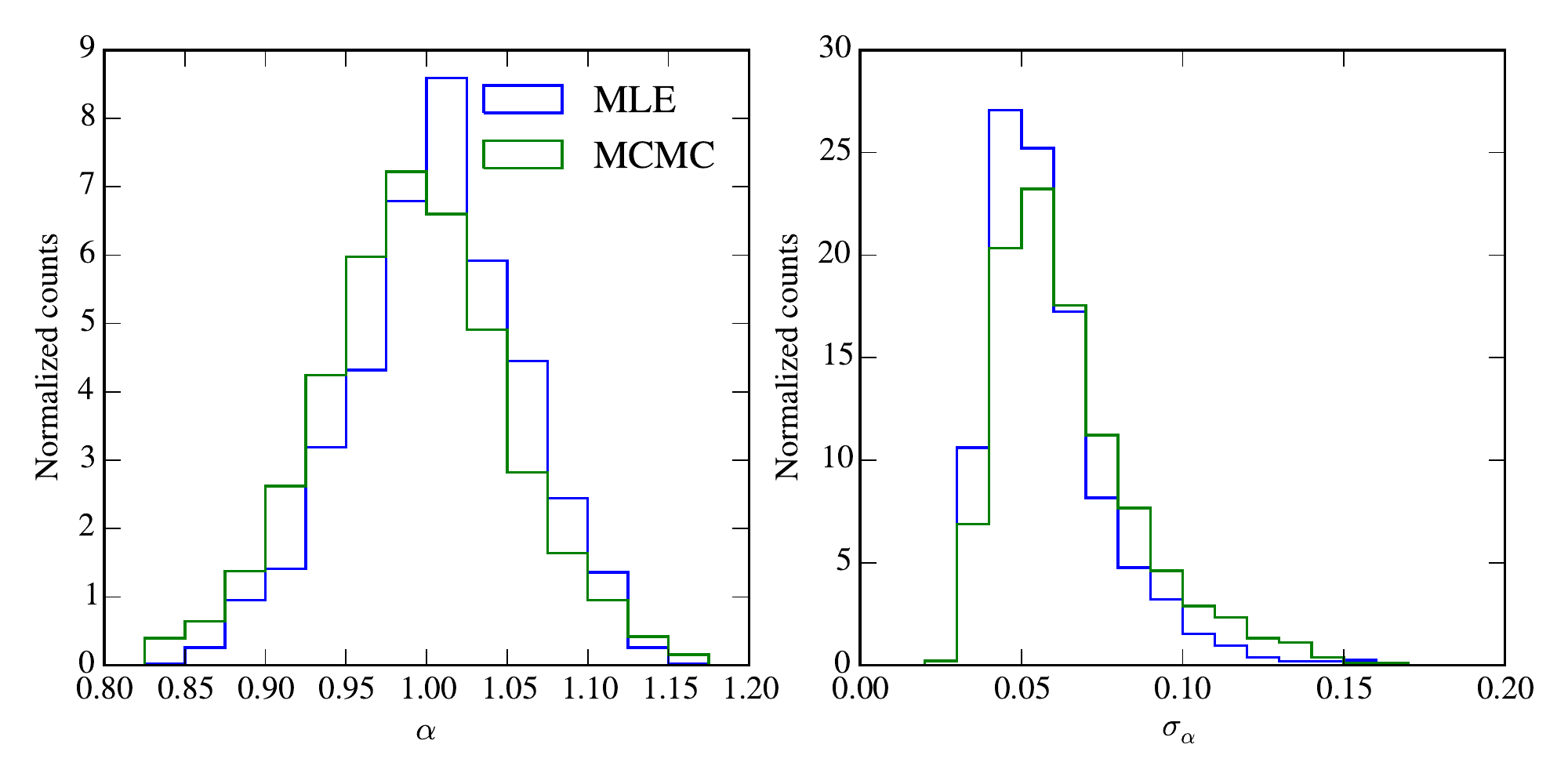}
    \caption{{\it Left}: distribution of the recovered $\alpha$ for the detected mocks.
      {\it Right}: distribution of the estimated error on $\alpha$.
      Results from different methods are presented. }
  \label{fig:alpha_dist}
\end{figure}

In Fig.~\ref{fig:alpha_dist} we show the distribution of $\alpha$ values resulting from fits  of our $C_\ell$ measurements in four photo-$z$ bins for 
the 1800 mocks. 
The remaining 16 parameters are marginalized over for the MCMC analysis and fixed to the values that maximize the  likelihood for the MLE analysis as described in \cite{2018arXiv180104390C}.
For the MLE method we find the best fit analytically over the 12 parameters $A_0, A_1, A_2$ in each redshift bin and numerically over the 4 $B_0$'s requiring $B_0 >0$ for each value of $\alpha$, following \cite{2018arXiv180104390C} . 

For the MCMC we used the flat priors $\alpha \in [0.8, 1.2]$, 
$A_1 \in [-800, 800] \times 10^{10}$, 
$A_2 \in [-50, 50] \times 10^{6}$, 
$A_3 \in [-200, 200] \times 10^{3}$ and 
$B_0 \in (0, 6]$.

We exclude outliers defined as mocks whose $1\sigma$ values for $\alpha$ lie outside the range $0.8 < \alpha < 1.2$  (see \cite{2018arXiv180104390C}) using the MLE method.
For our fiducial analysis $86.4 \%$ of the mocks are kept.
 
Since our fiducial model has the same cosmology as the mocks we expect to find $\alpha=1$. In fact we find that the mean value in the mocks is $\bar{\alpha}=1.006$ for MLE and $\bar{\alpha}=0.992$ for MCMC. Therefore both our methods recover $\alpha$ with a bias at the subpercent level.

\begin{figure}
\hspace*{-0.3cm}
\centering
    \includegraphics[width=\linewidth]{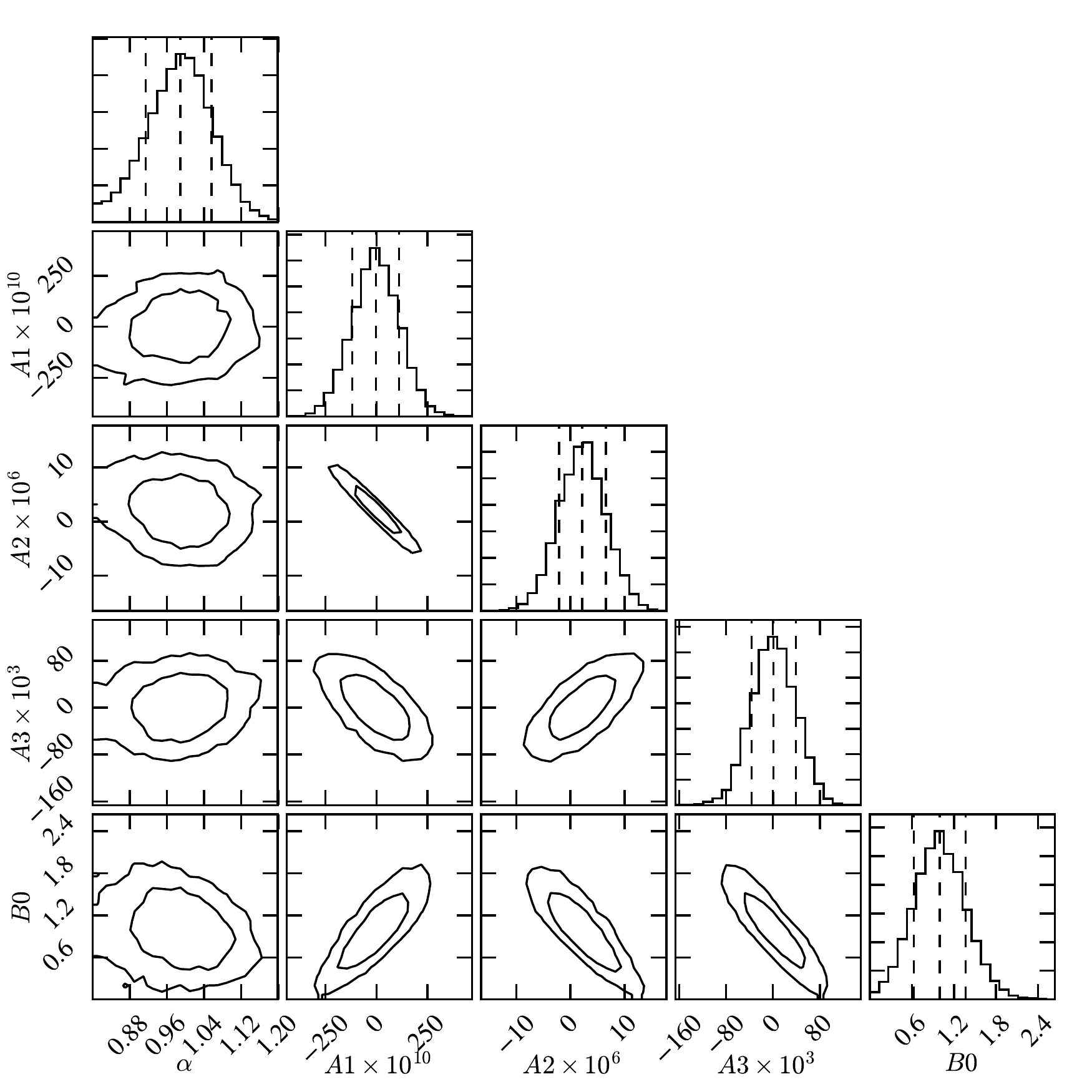}
  \caption{ Fit results for $\alpha$ and template parameters $B_0$, $A_0$, $A_1$ and $A_2$ 
  in the first photo-$z$ bin 
  for a BAO template fitted to the 
  average of the 1800 DES mock simulations. The plots for the parameters in other bins are similar. 
} 
  \label{fig:triangle}
\end{figure}

In Fig.~\ref{fig:triangle} we show the results from the MCMC chains when fitting 
the BAO template to the averaged $C_\ell$ measured  in the mock simulations  for our fiducial template. 
In this case, we find for the shift parameter $\alpha = 0.988 \pm 0.060$ and it can be seen that it does not show strong correlations with the {\it nuisance} parameters.
In fact, the {\it nuisance} parameters are poorly constrained, having broad distributions. The best-fit values for $B_0$ and $A_0$ are found to be roughly consistent with the squared bias and the shot-noise in each bin, respectively. 
For the MLE method we find $\alpha = 1.009 \pm 0.056$ from a fit to the average of the mocks.

\begin{figure}
\hspace*{-0.3cm}
\centering
    \includegraphics[width=\linewidth]{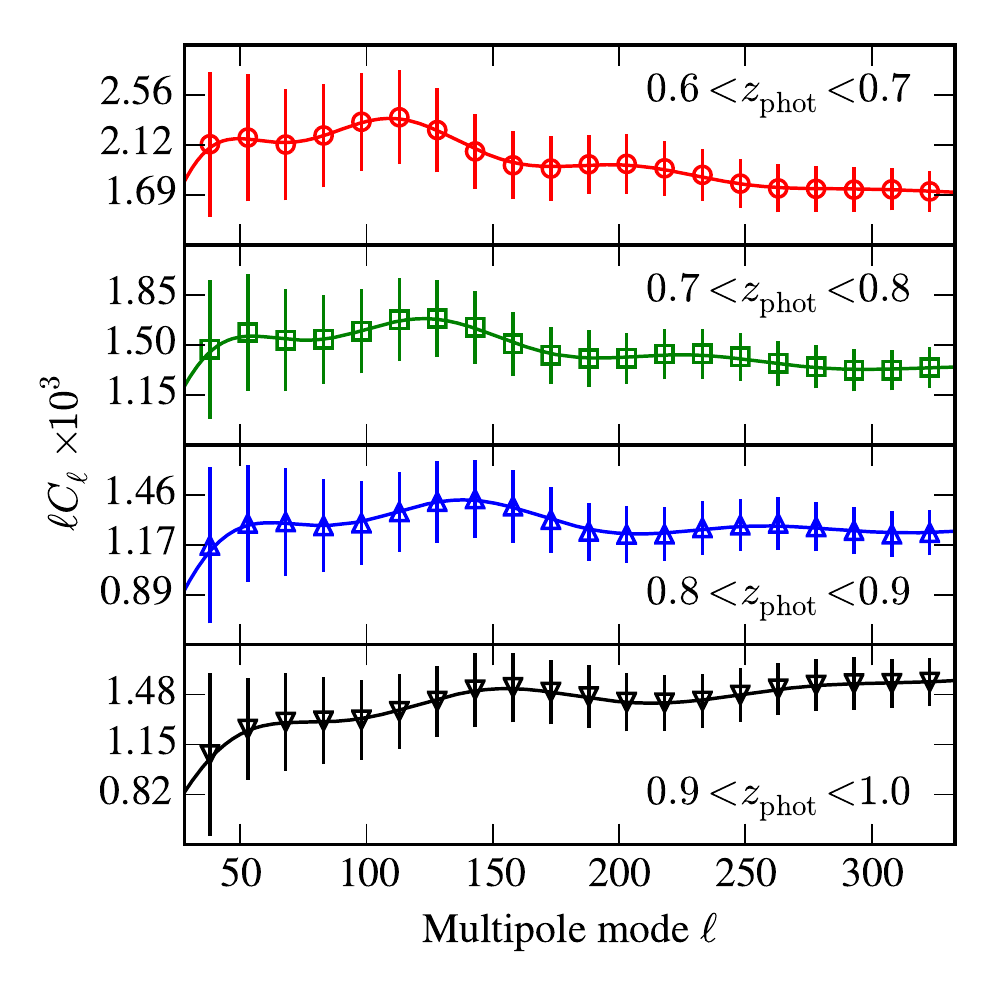}
  \caption{Measured $C_\ell$'s from DES mock simulations in four photo-$z$ bins. 
  The points show the average $C_\ell$'s from 1800 simulations and the error bars represent the diagonal of the covariance matrix of these measurements. 
  The line shows a theoretical prediction estimated at the simulation cosmology and best-fit template parameters.} 
  \label{fig:Cell}
\end{figure}

In Fig.~\ref{fig:Cell} we show $C_\ell$'s measured in four photo-$z$ bins for the DES mock simulations. 
The errors are computed from the mock covariance matrix.
The solid line displays the best-fit theoretical prediction using the BAO template described in \S~\ref{subsec:templates}.
We see that our BAO template is able to accurately capture the behaviour of the $C_\ell$'s from the mocks.  

\begin{figure}
\hspace*{-0.3cm}
\centering
    \includegraphics[width=\linewidth]{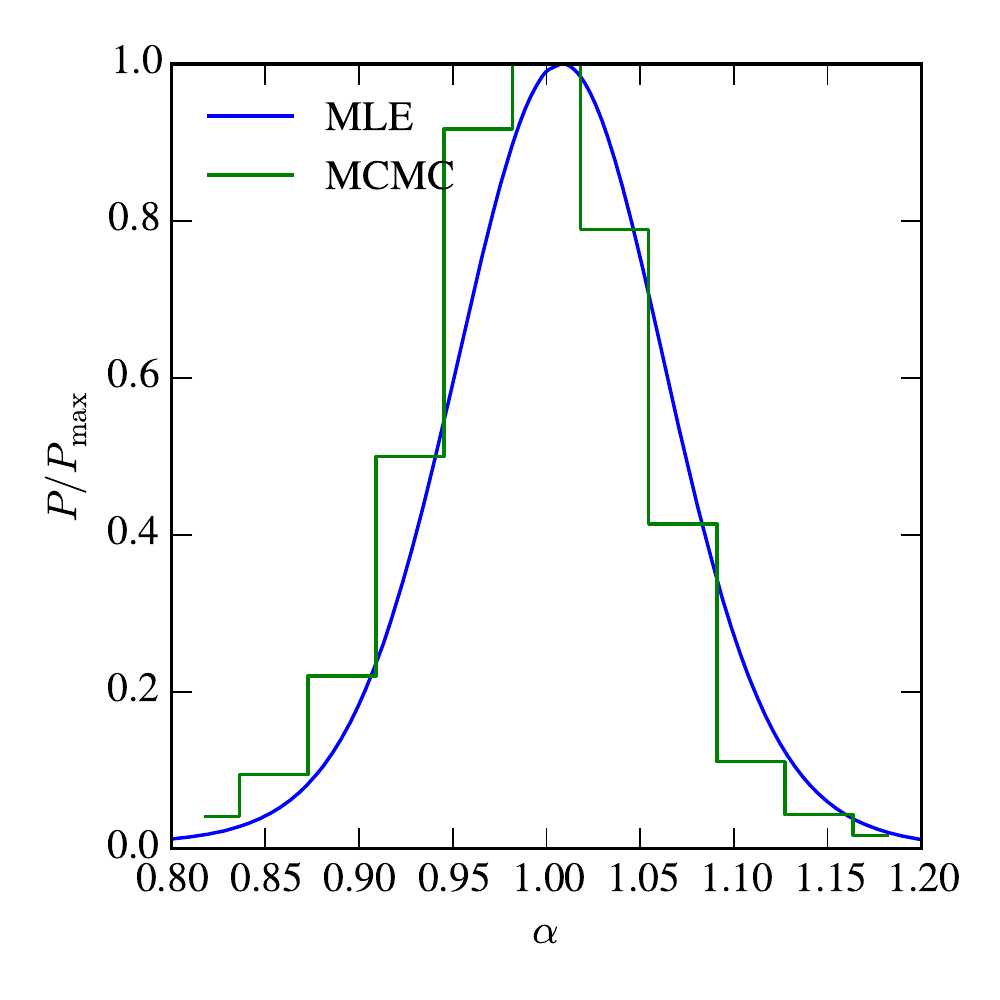}
  \caption{Normalized likelihoods from the MLE (solid line) and MCMC (histogram) methods for the $\alpha$ parameter determined from the average of the mock $C_\ell$'s.} 
  \label{fig:likelihoodsmocks}
\end{figure}

The compatibility between the two independent methods (MCMC and MLE) is shown in Fig.~\ref{fig:likelihoodsmocks} where we plot the normalized likelihood for the $\alpha$ parameter determined from the average of the mocks for both methods.

\begin{figure}
\hspace*{-0.3cm}
\centering
    \includegraphics[width=\linewidth]{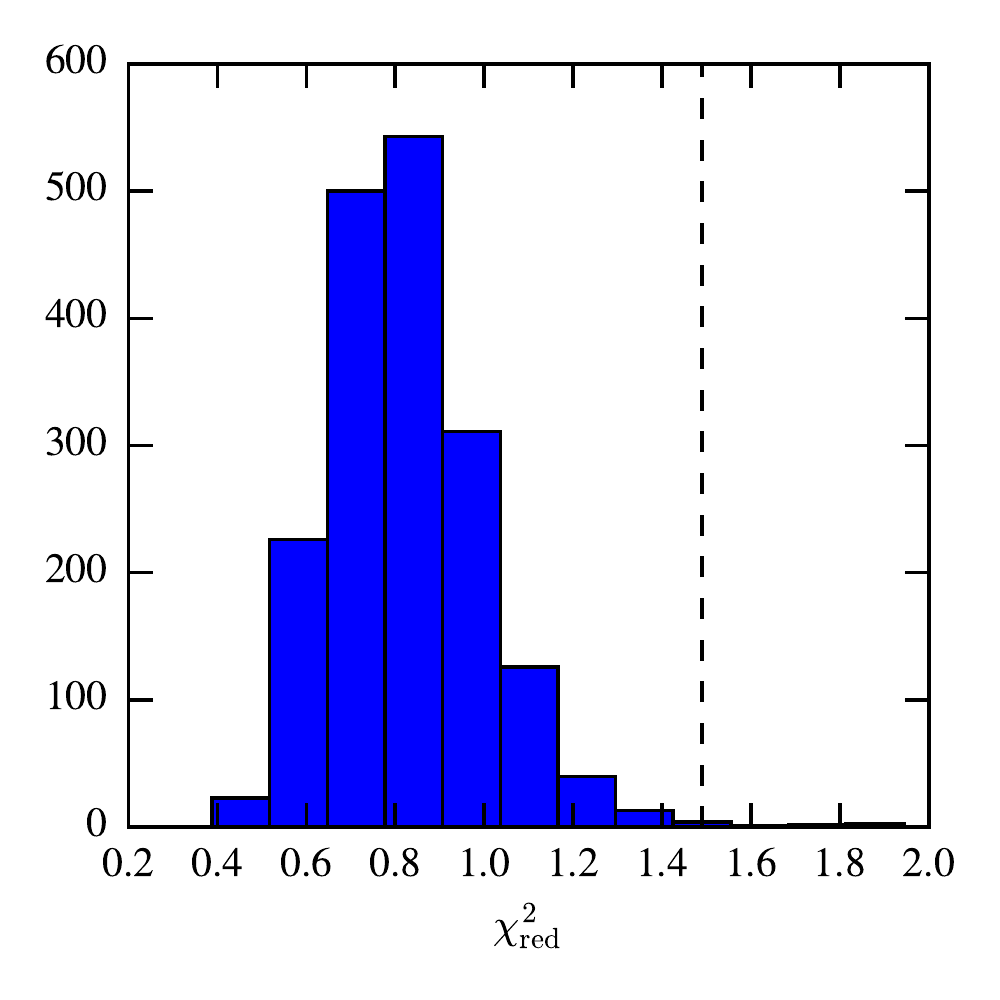}
  \caption{Distribution of the reduced $\chi^2$ values for the 1800 mocks. The dashed line shows the value of $\chi^2$ obtained from the data using the covariance matrix estimated from the mocks (\S~\ref{sec:BAOY1}).} 
  \label{fig:chi2mocks}
\end{figure}

In Fig.~\ref{fig:chi2mocks} we show the distribution of $\chi^2$ for the 1800 mocks demonstrating the good fit of our template. Also shown in the plot as a dashed line is the $\chi^2$ obtained from the data using the covariance matrix estimated from the mocks (discussed in \S~\ref{sec:BAOY1}).

\begin{figure}
\hspace*{-0.3cm}
\centering
    \includegraphics[width=\linewidth]{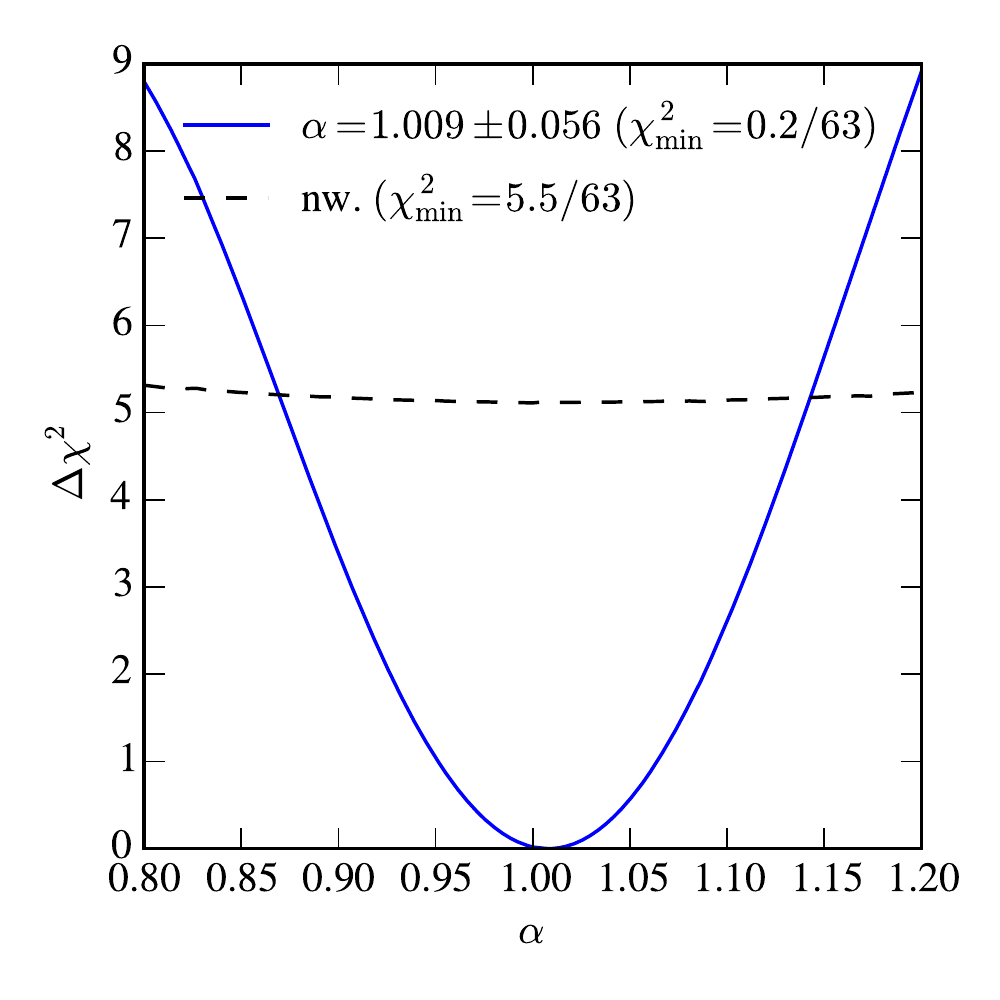}
    \caption{$\Delta \chi^2=\chi^2(\alpha) - \chi^2_{\rm min}$ as a function of $\alpha$ for the BAO template fitted on the mean of mocks. For each value of $\alpha$ we subtract from $\chi^2$ the value of $\chi^2_{\rm min}=\chi^2(\alpha_{min})$.
      Dashed line is the approximately constant $\chi^2$ for the non-wiggle template subtracted from the minimum of the template.} 
  \label{fig:deltachi2mocks}
\end{figure}

We estimate the significance of recovering $\alpha$ (or detecting the BAO feature) by measuring the difference in $\chi^2$ as a function of  $\alpha$ between a model with no BAO feature (a no-wiggle model), which is independent of $\alpha$, and our BAO template.
In Fig.~\ref{fig:deltachi2mocks} we show  $\Delta \chi^2=\chi^2(\alpha)-\chi^2_{\rm min}$ for fits of the average $C_\ell$'s from the mocks as a function of the $\alpha$ parameter. The best-fit value is $\alpha_{\rm min}=1.009$. 
From Fig.~\ref{fig:deltachi2mocks} we see that for the average of the mocks a BAO signal would be detected at $2.3 \sigma$ with respect to a no-wiggle model.

We will use the methods described in this section to study the BAO signal in Y1 data. But before doing so, we use the mocks to perform some robustness tests related to choices made in our analysis.

\subsection{Robustness Tests}
\label{subsec:rob}

For our default analysis above, a number of choices were made: the binning of harmonics in $\Delta\ell=15$, adopting $\ell_{\rm min} = 30$ and $\ell_{\rm max} = 330$, and the fiducial template used. We recall that we are including linear RSD in the modelling and we are using the full covariance matrix with redshift bin cross-correlations.  
We have examined the impact on the parameter estimation and on the fraction of detection of the mocks (the fraction of mocks remaining when excluding outliers) for some other choices. A summary of some of our tests is shown in Table~\ref{table:tests}, for choices of binning and range of $\ell$ as well as $C_\ell$ templates. We conclude that our choice of template gives an unbiased result for $\alpha$ at the percent level and a reasonable detection fraction. Although different choices produce small changes in the fits, they do not affect the BAO detection significantly, showing that our analysis is robust.

\begin{table*}
\centering
\begin{tabular}{lccccc}
\hline\hline
case & $\left\langle\alpha\right\rangle$ & $\left\langle\sigma\right\rangle$ & $S_\alpha$ & $f(N_{\rm det})$ & mean of mocks\\
\hline\hline
$\Delta\ell=15$, $30<\ell<330$ : & & & &\\
\hline
$A_0+A_1 \ell+A_2 \ell^{-1}$ & $1.003$ & $0.051$ & $0.058$ & $0.752$ & $1.008 \pm 0.056\, $\\
$\mathbf{A_0+A_1 \ell+A_2 \ell^{-2}}$ & $\mathbf{1.007}$ & $\mathbf{0.058}$ & $\mathbf{0.053}$ & $\mathbf{0.864}$ & $\mathbf{1.009 \pm 0.056}\, $\\
$A_0+A_1 \ell+A_2 \ell^{2}$ & $1.011$ & $0.056$ & $0.055$ & $0.851$ & $1.013 \pm 0.056\, $ \\
\hline
$\Delta\ell=20$, $40<\ell<300$ : & & & &\\
\hline
$A_0+A_1 \ell+A_2 \ell^{-1}$ & $1.003$ & $0.051$ & $0.060$ & $0.734$ & $1.006 \pm 0.058\, $\\
$A_0+A_1 \ell+A_2 \ell^{-2}$ & $1.006$ & $0.059$ & $0.056$ & $0.812$ & $1.006 \pm 0.058\, $\\
$A_0+A_1 \ell+A_2 \ell^{2}$ & $1.009$ & $0.057$ & $0.057$ & $0.790$ & $1.012 \pm 0.057\, $ \\
\hline
$\Delta\ell=15$, $45<\ell<330$ : & & & &\\
\hline
$A_0+A_1 \ell+A_2 \ell^{-1}$ & $1.004$ & $0.050$ & $0.059$ & $0.736$ & $1.009 \pm 0.056\, $\\
$A_0+A_1 \ell+A_2 \ell^{-2}$ & $1.007$ & $0.057$ & $0.054$ & $0.841$ & $1.009 \pm 0.056\, $\\
$A_0+A_1 \ell+A_2 \ell^{2}$ & $1.011$ & $0.056$ & $0.055$ & $0.839$ & $1.013 \pm 0.056\, $ \\
\hline
$\Delta\ell=20$, $40<\ell<320$ : & & & &\\
\hline
$A_0+A_1 \ell+A_2 \ell^{-1}$ & $1.004$ & $0.050$ & $0.060$ & $0.731$ & $1.008 \pm 0.056\, $\\
$A_0+A_1 \ell+A_2 \ell^{-2}$ & $1.007$ & $0.058$ & $0.055$ & $0.833$ & $1.008 \pm 0.057\, $\\
$A_0+A_1 \ell+A_2 \ell^{2}$ & $1.011$ & $0.056$ & $0.057$ & $0.831$ & $1.014 \pm 0.057\,$ \\
\hline
\end{tabular}
\caption{Summary of the robustness tests performed on the 1800 mocks using MLE. We show the average values of $\alpha$ and its $1\sigma$ standard deviation for all the mocks, the standard deviation of $\alpha$ obtained only for the detected mocks $S_\alpha$ and the fraction of detected mocks.
The fiducial case we adopt has a template $A_0+A_1 \ell+A_2 \ell^{-2}$ and  $\Delta\ell=15$, $30 < \ell < 330$ shown in boldface.}
\label{table:tests}
\end{table*}

In addition to the tests in Table~\ref{table:tests}, we have also investigated other choices made. These included i) using the Limber approximation \citep{1953ApJ...117..134L} instead of the full integral calculation in Eqs.~(\ref{eq:Cell_theory}) and (\ref{eq:Psi_RSD}), ii) exclusion of linear RSD effects, i.e. using Eq.~(\ref{eq:Psi}) instead of Eq.~(\ref{eq:Psi_RSD}), iii) exclusion of cross-correlations between photo-$z$ bins in the covariance matrix, iv) use of the theoretical covariance instead of the covariance measured in mocks and v) inclusion of the non-linear matter power spectrum in the $C_\ell$ modeling.  All these tests led to very similar results as the fiducial analysis for $\alpha$. 

Notice that i) and v) affect only small scales, ii) affects only large scales. Meanwhile we expect iii) and iv) to have small effects given that redshift cross-correlations are small for our photo-$z$ bin size and the theoretical covariance matches that measured in the mocks quite well (see \S~\ref{subsec:cov}). Our $C_\ell$ template has enough flexibility to account for these effects in case they are either included or neglected. Indeed we find that the best-fit template parameters change significantly between one case and another, but the best-fit for $\alpha$ and the BAO detection significance remain nearly the same.

\section{BAO in DES Y1 data}\label{sec:BAOY1}

\begin{figure}
\hspace*{-0.3cm}
\centering
    \includegraphics[width=\linewidth]{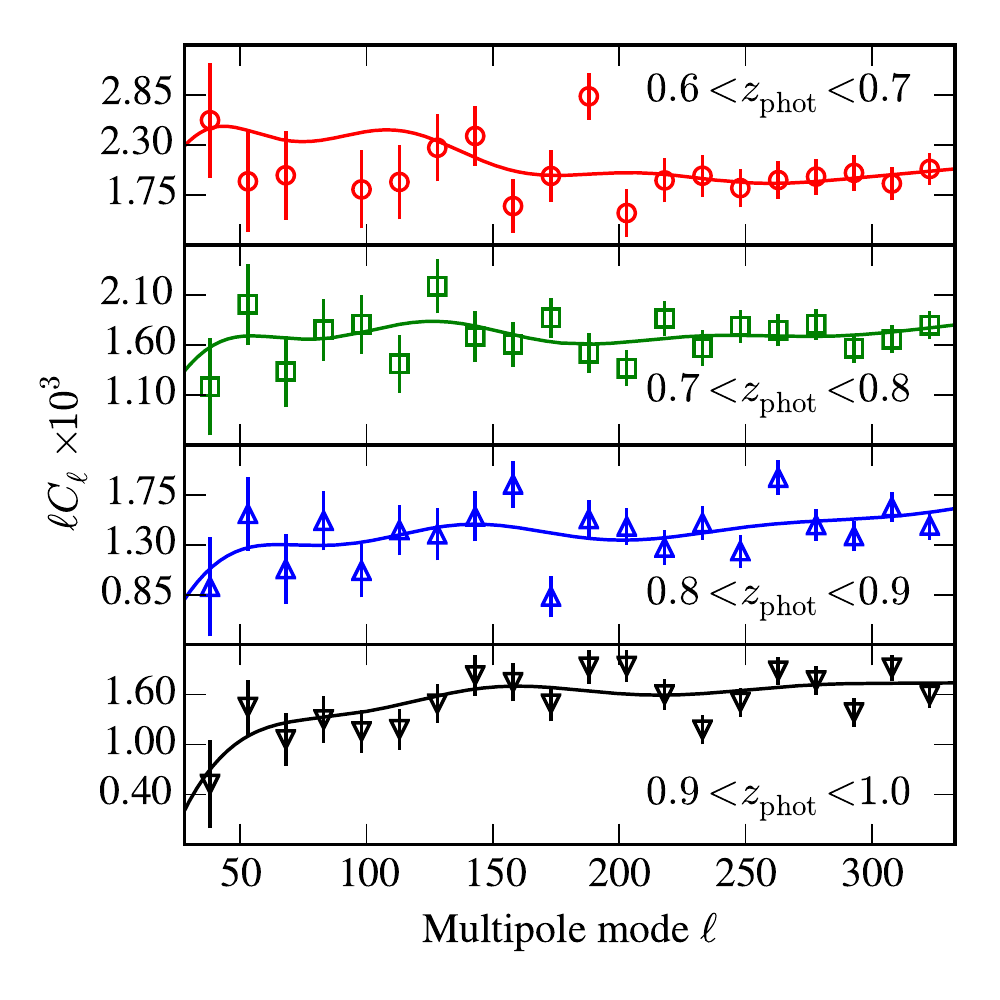}
  \caption{Measured $C_\ell$'s from DES Y1 data in four photo-$z$ bins. 
  The errors represent the diagonal of the covariance of 1800 mock simulations. 
  The line shows our best fits from the fiducial analysis.} 
  \label{fig:Celly1}
\end{figure}

We now apply the methods described and tested above to study the BAO feature in the angular power spectrum in the DES  Y1 data.
In Fig.~\ref{fig:Celly1} we show $C_\ell$'s measured in four photo-$z$ bins for the DES Y1 data. The errors are computed from the variance of the 1800 mock simulations. The solid line displays the best-fit theoretical prediction using the BAO template described in \S~\ref{sec:theory}.  

\begin{figure}
\hspace*{-0.3cm}
\centering
    \includegraphics[width=\linewidth]{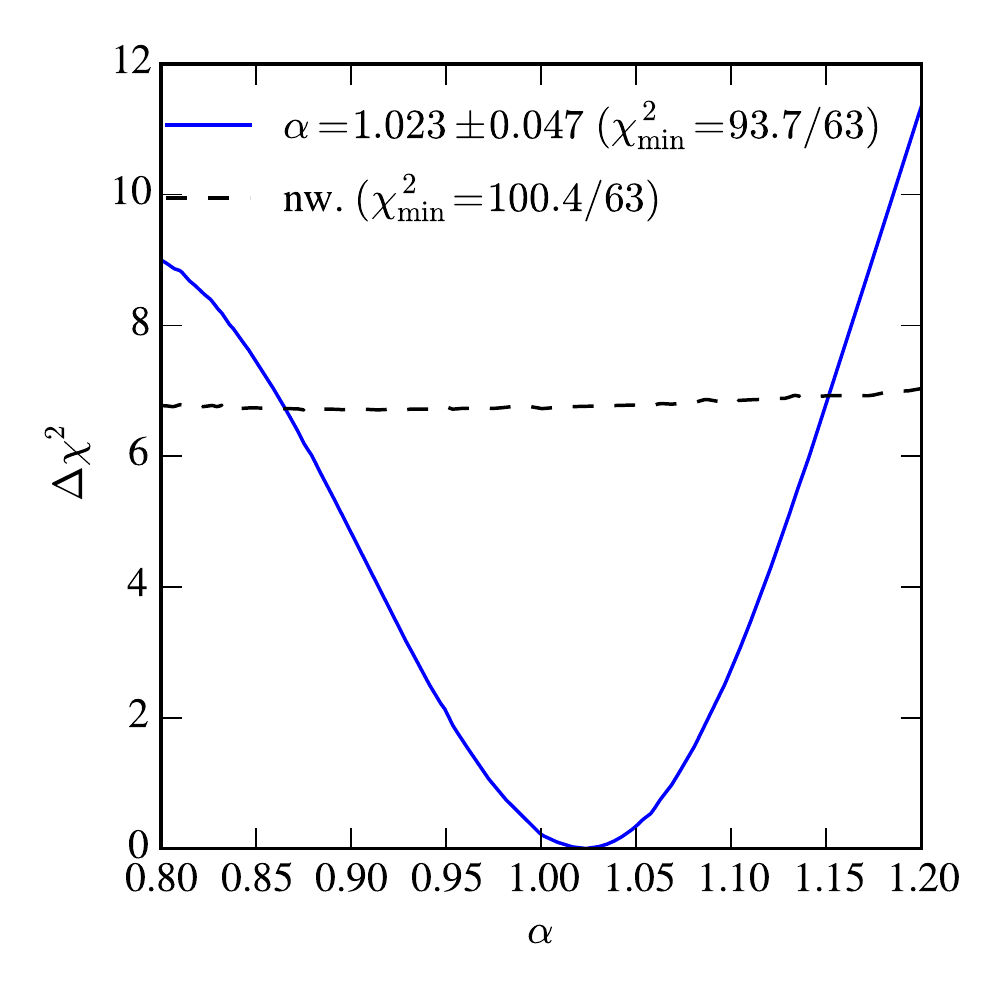}
  \caption{$\Delta \chi^2$ as a function of $\alpha$ for the DES Y1 galaxy data, when fitted to a BAO templates (solid blue curve) and to a no-wiggle template (dashed black curve).} 
  \label{fig:deltachi2y1data}
\end{figure}

In Fig.~\ref{fig:deltachi2y1data} we show the $\Delta \chi^2=\chi^2(\alpha)-\chi^2_{\rm min}$ of the fits as a function of $\alpha$ for the DES Y1 data from the MLE described above and also used in \cite{2018arXiv180104390C}. We find $\alpha=1.023 \pm 0.047$ with $\chi^2_{\rm min}/{\rm dof}=93.7/63=1.49$. This somewhat large value of $\chi^2$ seems to  indicate that the covariance matrix obtained from the mocks may underestimate the errors. We will discuss this possibility below.

The small deviation of $\alpha$ from unity can be traced to the fact that the template cosmology has been fixed to reflect that of the mock simulations (to be consistent with the fact that we also use the covariance from the mock simulations). The mocks have a cosmology slightly different from e.g. the Planck cosmology, and the latter has been shown to be consistent with clustering measurements of the DES Y1 data \citep{Gruen:2017xjj,Abbott:2017wau}.
A difference of a few percent in $\alpha$ from unity is expected and is also found in a similar analysis in configuration space \citep{2018arXiv180104390C,Abbott:2017wcz}. 
We have repeated our analysis with the covariance matrix re-calculated at the
best fit cosmology, and we have not found significant changes in our results, which was also the case for \citet{Abbott:2017wau}.

Finally, we also show the difference in $\chi^2$ from our best-fit template and a no-wiggle model.
Assuming Gaussian statistics for the likelihood, we find that the angular power spectrum measured from DES Y1 data finds the BAO feature at a significance of $2.6 \,\sigma$ level with respect to a no-wiggle template.

In order to address the issue of the large value of $\chi^2$ obtained above we study the changes that arise from using a theoretical covariance matrix more well adjusted to the data. We modify the theoretical modelling of $C_\ell$ used to fit the average of the mocks in Fig.~\ref{fig:MocksDataTheory} by adding a term proportional to $\ell$ and fit its coefficient to best reproduce the data. This theoretical $C_\ell$ is then input in {\tt NaMaster} to compute a new gaussian covariance matrix that takes into account the Y1 mask and the binning in $\ell$. When this new covariance matrix is used the minimum $\chi^2$ is indeed reduced to $\chi^2_{\rm min}/{\rm dof}=85.8/63=1.36$ without a significant change in the estimated value of $\alpha$, 
which is found to be $\alpha=1.039 \pm 0.053$ in this case.

\begin{figure}
\hspace*{-0.3cm}
\centering
    \includegraphics[width=\linewidth]{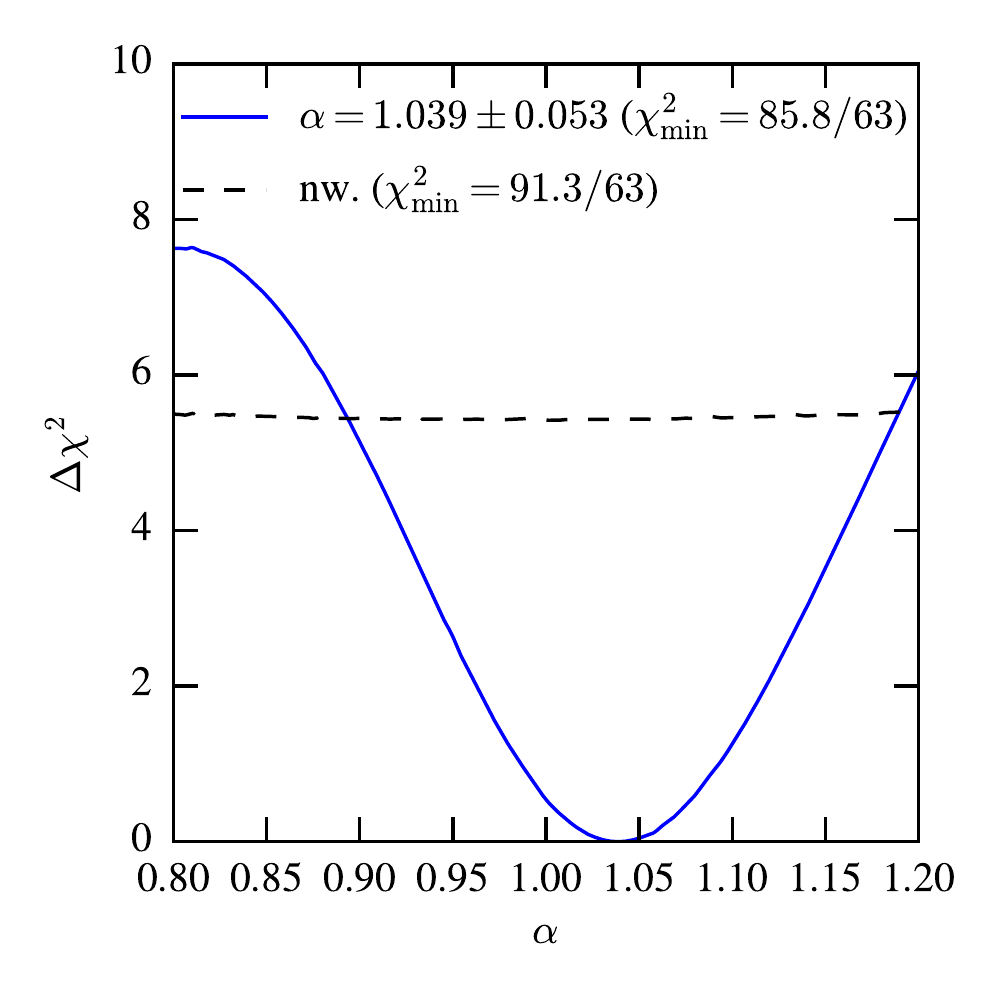}
  \caption{$\Delta \chi^2$ as a function of $\alpha$ for the DES Y1 galaxy data, when fitted using a new theoretical covariance matrix to a BAO template (solid blue curve) and to a no-wiggle template (dashed black curve).} 
  \label{fig:deltachi2y1data_new}
\end{figure}

In Fig.~\ref{fig:deltachi2y1data_new} we show the result of the significance using this new theoretical covariance matrix. The value of $\alpha$ changed by a third of the standard deviation and the error increased by $13\%$. Although the changes are small they point to the uncertainties inherent in this analysis.

\section{Conclusions}\label{sec:conclusions}

The DES is on its way to produce the largest survey to date, projected to map $300$ million galaxies using photometric techniques in an area of 5000 deg$^2$ up to a redshift $z \approx 1.3$. The Y1 data has been recently analysed resulting in a key paper combining three correlations: weak gravitational lensing, galaxy clustering and their cross-correlation or galaxy-galaxy lensing \citep{Abbott:2017wau}. 
Several papers dealing with the essential developments that led to the key paper were also produced \citep{Cawthon:2017qxi,Avila:2017nyy,Davis:2017dlg,Gatti:2017hmb,Hoyle:2017mee,Drlica-Wagner:2017tkk,Krause:2017ekm}.

The work presented here is part of a series of papers mentioned in the Introduction dedicated to searching specifically for the BAO feature in Y1 data. Here we concentrated on the detection of the BAO feature in the angular power spectrum.   

We developed a methodology based on template-fitting and tested it on realistic DES Y1 galaxy mocks. 
First we tested two independent codes for pseudo-$C_\ell$ estimators  and found agreement between codes to better than 1\% in nearly all scales of interest. We then measured the APS in four photo-$z$ bins for 1800 mock catalogs, checking their consistency with theoretical expectations. 
We measured the covariance matrix from the mocks and compared it with a theoretical prediction, finding again good agreement.
We then used two independent methods, a maximum likelihood estimator (MLE) and a Markov Chain Monte Carlo (MCMC) analysis to estimate the shift parameter for the average of the mocks and found the two methods to be compatible. Comparing the values of $\chi^2$ for our BAO template to a no-wiggle model we find a $2.3 \,\sigma$ signal for BAO in the mocks.
 
Several choices were made for our fiducial analysis and we performed a number of robustness tests to assess their impact on the results. 
We find that our results on the mocks were not significantly sensitive to changing the binning $\Delta \ell$ from $10$ to $20$, changing the smallest scales of our analysis from  $\ell_{\rm max} = 300$ to $\ell_{\rm max} = 330$, neglecting the redshift cross-covariance, using the Limber approximation, neglecting linear RSD's, including a non-linear matter power spectrum and modifying the $C_\ell$ template used.

We then applied the fiducial analysis to measure the APS of a galaxy sample obtained from DES Y1 data also split into four photo-$z$ bins up to $z = 1$ \citep{Crocce:2017iwq}. 
We obtain a best-fit $\alpha=1.023 \pm 0.047$. This corresponds to a measurement of the ratio of the angular diameter distance to the effective redshift of our sample ($z_{\rm eff} = 0.81$) and the BAO physical scale $r_d$ of $D_A(z_{\rm eff} = 0.81)/r_d = 10.65 \pm 0.49$. Comparing to the best-fit no-wiggle template we find a significance of $2.6 \,\sigma$ for BAO detection.

This best fit has a somewhat large $\chi^2$/(dof)=1.49 value. We could trace the reason to the covariance matrix computed from the mocks, since the $C_\ell$'s measured from them seem to underestimate the data at high $\ell$'s in two redshift bins. We investigate this issue with a new Gaussian theoretical covariance matrix obtained from $C_\ell$'s that are better adjusted to the data, taking into account the mask and the binning. With this new covariance matrix we obtain a reduced value $\chi^2=1.36$ without significant changes in the recovered value of $\alpha$.

Our results are consistent with those from the real-space BAO analysis of Y1 data \citep{Abbott:2017wau} but the methodological uncertainties we found, despite being small, must be understood in more detail in future DES analyses.

The use of photometric data such as that from DES allows us to extend the BAO detection to high-redshift galaxies.
The consistency of the BAO scale inferred from CMB and galaxies is an important test of the standard cosmological model over most of the cosmic history. 
As DES continues to collect and analyze more data, the significance of the BAO feature detection will continue to improve. Data collected over three years of 
observations (Y3) covers nearly the whole DES footprint.
In combination with additional probes of geometry and structure growth, the BAO feature detected in this extended area of DES will be an important element for constraining and distinguishing models of cosmic acceleration in the near future. 

\section*{Acknowledgments}
HC is supported by CNPq under grant number 141935/2014-6. 
ML and RR are partially supported by FAPESP and CNPq. 
AT is supported by FAPESP. 
We thank the support of the Instituto Nacional de Ci\^encia
e Tecnologia (INCT) e-Universe (CNPq grant 465376/2014-2).

We are grateful for the extraordinary contributions of our CTIO colleagues and the DECam Construction, Commissioning and Science Verification
teams in achieving the excellent instrument and telescope conditions that have made this work possible.  The success of this project also 
relies critically on the expertise and dedication of the DES Data Management group.

Funding for the DES Projects has been provided by the U.S. Department of Energy, the U.S. National Science Foundation, the Ministry of Science and Education of Spain, 
the Science and Technology Facilities Council of the United Kingdom, the Higher Education Funding Council for England, the National Center for Supercomputing 
Applications at the University of Illinois at Urbana-Champaign, the Kavli Institute of Cosmological Physics at the University of Chicago, 
the Center for Cosmology and Astro-Particle Physics at the Ohio State University,
the Mitchell Institute for Fundamental Physics and Astronomy at Texas A\&M University, Financiadora de Estudos e Projetos, 
Funda{\c c}{\~a}o Carlos Chagas Filho de Amparo {\`a} Pesquisa do Estado do Rio de Janeiro, Conselho Nacional de Desenvolvimento Cient{\'i}fico e Tecnol{\'o}gico and 
the Minist{\'e}rio da Ci{\^e}ncia, Tecnologia e Inova{\c c}{\~a}o, the Deutsche Forschungsgemeinschaft and the Collaborating Institutions in the Dark Energy Survey. 

The Collaborating Institutions are Argonne National Laboratory, the University of California at Santa Cruz, the University of Cambridge, Centro de Investigaciones Energ{\'e}ticas, 
Medioambientales y Tecnol{\'o}gicas-Madrid, the University of Chicago, University College London, the DES-Brazil Consortium, the University of Edinburgh, 
the Eidgen{\"o}ssische Technische Hochschule (ETH) Z{\"u}rich, 
Fermi National Accelerator Laboratory, the University of Illinois at Urbana-Champaign, the Institut de Ci{\`e}ncies de l'Espai (IEEC/CSIC), 
the Institut de F{\'i}sica d'Altes Energies, Lawrence Berkeley National Laboratory, the Ludwig-Maximilians Universit{\"a}t M{\"u}nchen and the associated Excellence Cluster Universe, 
the University of Michigan, the National Optical Astronomy Observatory, the University of Nottingham, The Ohio State University, the University of Pennsylvania, the University of Portsmouth, 
SLAC National Accelerator Laboratory, Stanford University, the University of Sussex, Texas A\&M University, and the OzDES Membership Consortium.

Based in part on observations at Cerro Tololo Inter-American Observatory, National Optical Astronomy Observatory, which is operated by the Association of 
Universities for Research in Astronomy (AURA) under a cooperative agreement with the National Science Foundation.

The DES data management system is supported by the National Science Foundation under Grant Numbers AST-1138766 and AST-1536171.
The DES participants from Spanish institutions are partially supported by MINECO under grants AYA2015-71825, ESP2015-66861, FPA2015-68048, SEV-2016-0588, SEV-2016-0597, and MDM-2015-0509, 
some of which include ERDF funds from the European Union. IFAE is partially funded by the CERCA program of the Generalitat de Catalunya.
Research leading to these results has received funding from the European Research
Council under the European Union's Seventh Framework Program (FP7/2007-2013) including ERC grant agreements 240672, 291329, and 306478.
We  acknowledge support from the Australian Research Council Centre of Excellence for All-sky Astrophysics (CAASTRO), through project number CE110001020.

This manuscript has been authored by Fermi Research Alliance, LLC under Contract No. DE-AC02-07CH11359 with the U.S. Department of Energy, Office of Science, Office of High Energy Physics. The United States Government retains and the publisher, by accepting the article for publication, acknowledges that the United States Government retains a non-exclusive, paid-up, irrevocable, world-wide license to publish or reproduce the published form of this manuscript, or allow others to do so, for United States Government purposes.

This paper has gone through internal review by the DES collaboration.
The DES publication number for this article is DES-2017-0307. 

\bibliographystyle{mnras}
\bibliography{APS}

\section*{Affiliations}
$^{1}$ Departamento de F\'isica Matem\'atica, Instituto de F\'isica, Universidade de S\~ao Paulo, CP 66318, S\~ao Paulo, SP, 05314-970, Brazil\\
$^{2}$ Laborat\'orio Interinstitucional de e-Astronomia - LIneA, Rua Gal. Jos\'e Cristino 77, Rio de Janeiro, RJ - 20921-400, Brazil\\
$^{3}$ Instituto de F\'{\i}sica Te\'orica, Universidade Estadual Paulista, S\~ao Paulo, Brazil\\
$^{4}$ ICTP South American Institute for Fundamental Research\\ Instituto de F\'{\i}sica Te\'orica, Universidade Estadual Paulista, S\~ao Paulo, Brazil\\
$^{5}$ D\'{e}partement de Physique Th\'{e}orique and Center for Astroparticle Physics, Universit\'{e} de Gen\`{e}ve, 24 quai Ernest Ansermet, CH-1211 Geneva, Switzerland\\
$^{6}$ Instituto de F\'isica Gleb Wataghin, Universidade Estadual de Campinas, 13083-859, Campinas, SP, Brazil\\
$^{7}$ Observat\'orio Nacional, Rua Gal. Jos\'e Cristino 77, Rio de Janeiro, RJ - 20921-400, Brazil\\
$^{8}$ Institute of Cosmology \& Gravitation, University of Portsmouth, Portsmouth, PO1 3FX, UK\\
$^{9}$ Institut d'Estudis Espacials de Catalunya (IEEC), 08193 Barcelona, Spain\\
$^{10}$ Institute of Space Sciences (ICE, CSIC),  Campus UAB, Carrer de Can Magrans, s/n,  08193 Barcelona, Spain\\
$^{11}$ Center for Cosmology and Astro-Particle Physics, The Ohio State University, Columbus, OH 43210, USA\\
$^{12}$ Instituto de Fisica Teorica UAM/CSIC, Universidad Autonoma de Madrid, 28049 Madrid, Spain\\
$^{13}$ Cerro Tololo Inter-American Observatory, National Optical Astronomy Observatory, Casilla 603, La Serena, Chile\\
$^{14}$ Department of Physics \& Astronomy, University College London, Gower Street, London, WC1E 6BT, UK\\
$^{15}$ Department of Physics and Electronics, Rhodes University, PO Box 94, Grahamstown, 6140, South Africa\\
$^{16}$ Fermi National Accelerator Laboratory, P. O. Box 500, Batavia, IL 60510, USA\\
$^{17}$ Observatories of the Carnegie Institution of Washington, 813 Santa Barbara St., Pasadena, CA 91101, USA\\
$^{18}$ CNRS, UMR 7095, Institut d'Astrophysique de Paris, F-75014, Paris, France\\
$^{19}$ Sorbonne Universit\'es, UPMC Univ Paris 06, UMR 7095, Institut d'Astrophysique de Paris, F-75014, Paris, France\\
$^{20}$ Jodrell Bank Center for Astrophysics, School of Physics and Astronomy, University of Manchester, Oxford Road, Manchester, M13 9PL, UK\\
$^{21}$ Kavli Institute for Particle Astrophysics \& Cosmology, P. O. Box 2450, Stanford University, Stanford, CA 94305, USA\\
$^{22}$ SLAC National Accelerator Laboratory, Menlo Park, CA 94025, USA\\
$^{23}$ Department of Astronomy, University of Illinois at Urbana-Champaign, 1002 W. Green Street, Urbana, IL 61801, USA\\
$^{24}$ National Center for Supercomputing Applications, 1205 West Clark St., Urbana, IL 61801, USA\\
$^{25}$ Institut de F\'{\i}sica d'Altes Energies (IFAE), The Barcelona Institute of Science and Technology, Campus UAB, 08193 Bellaterra (Barcelona) Spain\\
$^{26}$ Kavli Institute for Cosmological Physics, University of Chicago, Chicago, IL 60637, USA\\
$^{27}$ Department of Physics and Astronomy, University of Pennsylvania, Philadelphia, PA 19104, USA\\
$^{28}$ Centro de Investigaciones Energ\'eticas, Medioambientales y Tecnol\'ogicas (CIEMAT), Madrid, Spain\\
$^{29}$ Department of Physics, IIT Hyderabad, Kandi, Telangana 502285, India\\
$^{30}$ Department of Astronomy, University of Michigan, Ann Arbor, MI 48109, USA\\
$^{31}$ Department of Physics, University of Michigan, Ann Arbor, MI 48109, USA\\
$^{32}$ Institute of Astronomy, University of Cambridge, Madingley Road, Cambridge CB3 0HA, UK\\
$^{33}$ Kavli Institute for Cosmology, University of Cambridge, Madingley Road, Cambridge CB3 0HA, UK\\
$^{34}$ Universit\"ats-Sternwarte, Fakult\"at f\"ur Physik, Ludwig-Maximilians Universit\"at M\"unchen, Scheinerstr. 1, 81679 M\"unchen, Germany\\
$^{35}$ Santa Cruz Institute for Particle Physics, Santa Cruz, CA 95064, USA\\
$^{36}$ Department of Physics, The Ohio State University, Columbus, OH 43210, USA\\
$^{37}$ Max Planck Institute for Extraterrestrial Physics, Giessenbachstrasse, 85748 Garching, Germany\\
$^{38}$ Harvard-Smithsonian Center for Astrophysics, Cambridge, MA 02138, USA\\
$^{39}$ Department of Astronomy/Steward Observatory, 933 North Cherry Avenue, Tucson, AZ 85721-0065, USA\\
$^{40}$ Jet Propulsion Laboratory, California Institute of Technology, 4800 Oak Grove Dr., Pasadena, CA 91109, USA\\
$^{41}$ Australian Astronomical Observatory, North Ryde, NSW 2113, Australia\\
$^{42}$ George P. and Cynthia Woods Mitchell Institute for Fundamental Physics and Astronomy, and Department of Physics and Astronomy, Texas A\&M University, College Station, TX 77843,  USA\\
$^{43}$ Instituci\'o Catalana de Recerca i Estudis Avan\c{c}ats, E-08010 Barcelona, Spain\\
$^{44}$ Department of Physics and Astronomy, University of Waterloo, 200 University Ave W, Waterloo, ON N2L 3G1, Canada\\
$^{45}$ Perimeter Institute for Theoretical Physics, 31 Caroline St. North, Waterloo, ON N2L 2Y5, Canada\\
$^{46}$ Department of Physics and Astronomy, Pevensey Building, University of Sussex, Brighton, BN1 9QH, UK\\
$^{47}$ School of Physics and Astronomy, University of Southampton,  Southampton, SO17 1BJ, UK\\
$^{48}$ Brandeis University, Physics Department, 415 South Street, Waltham MA 02453\\
$^{49}$ Computer Science and Mathematics Division, Oak Ridge National Laboratory, Oak Ridge, TN 37831\\
$^{50}$ Institute for Astronomy, University of Edinburgh, Edinburgh EH9 3HJ, UK

\renewcommand{\thesubsection}{\Alph{subsection}}

\bsp%
\label{lastpage}

\end{document}